\documentclass[fleqn,usenatbib]{mnras}

\usepackage[T1]{fontenc}

\DeclareRobustCommand{\VAN}[3]{#2}
\let\VANthebibliography\thebibliography
\def\thebibliography{\DeclareRobustCommand{\VAN}[3]{##3}\VANthebibliography}

\usepackage{graphicx}	
\usepackage{amsmath}	
\usepackage{amstext,amscd,bm}
\usepackage{array}
\usepackage{orcidlink}

\title[Revisiting the Double Cluster in Perseus]{Revisiting the Double Cluster in Perseus: Are NGC 869 and NGC 884 a true binary system?}

\author[Zahra Al et al.]{
	Zahra Al\thanks{E-mail: zahraa96al@gmail.com}$^{1}$ \orcidlink{0009-0004-4134-3752},
	Juan Casado$^2$ \orcidlink{0000-0003-4105-2520},
	Sergio Sánchez-Sanjuán$^3$ \orcidlink{0000-0002-2269-9348}, 	
	Y\"uksel Karata\c{s}$^{4}$ \orcidlink{0000-0002-7737-6589},
	Yasser Hendy$^5$ \orcidlink{0000-0002-2356-8315},
    \newauthor
	Ra\'ul Michel$^3$ \orcidlink{0000-0003-1263-808X},
	Dana Kovaleva$^6$ \orcidlink{0000-0003-4328-8801},
	Doaa Eid$^5$ \orcidlink{0000-0003-0948-1755} 
	\\
	$^{1}$ Institute of Graduate Studies In Sciences, {\.I}stanbul University, 34116, \"Universite-Istanbul, T\"urkiye\\
	$^{2}$ Facultad de Ciencias, Universitat Autònoma de Barcelona (UAB), 08193 Bellaterra, Barcelona, Spain \\
    $^{3}$ Universidad Nacional Autónoma de México. Instituto de Astronomía. A.P. 106, 22800. Ensenada, B.C., México\\
	$^{4}$ Department of Astronomy and Space Sciences, Science Faculty, {\.I}stanbul University, 34116, \"Universite-Istanbul, T\"urkiye\\
	$^{5}$ Astronomy Department, National Research Institute of Astronomy and Geophysics (NRIAG), 11421 Helwan, Cairo, Egypt\\ 
	$^{6}$ Institute of Astronomy, Russian Academy of Sciences, 48 Pyatnitskaya Str., 119017, Moscow, Russian Federation\\}

\date{Accepted XXX. Received YYY; in original form ZZZ}

\pubyear{2026}

\begin{document}
\label{firstpage}
\pagerange{\pageref{firstpage}--\pageref{lastpage}}
\maketitle
	
\begin{abstract}
    
We perform a comprehensive study of the Double Cluster in Perseus, NGC 869 and NGC 884, to investigate their nature and evolutionary state. By combining new CCD $UBV(RI)_{KC}$ photometry with high-precision astrometry and photometry from \textit{Gaia} DR3, we determine robust astrophysical parameters, including colour excess, distance, and age. Our analysis confirms that both clusters are young, with an age of $12\pm1$ Myr, and share similar photometric distances of $2.29$ kpc. By integrating these observational constraints with completeness-corrected mass estimates and a suite of N-body simulations, we examine the kinematic and phase-space properties of the system. We find that the internal kinematics of NGC 884 exhibit a stronger and more coherent expansion signature than NGC 869. Our dynamical simulations demonstrate a rapid increase in the separation of the pair during the first 100 Myr, exceeding $1$ kpc. These results indicate that the system is not currently gravitationally bound and does not constitute a binary cluster.
Instead, the Double Cluster is presently unbound, and the forward integrations predict that its components will continue to diverge under the influence of their relative motion and the Galactic potential.


\end{abstract}
	
	
	
	
\section{Introduction}

The double cluster formed by NGC 869 and NGC 884 (h and $\chi$ Persei) is one of the most famous and well-studied open cluster (OC) systems in the Galaxy and the only known double cluster before the \textit{Gaia} era. Located between the Perseus and Cassiopeia constellations and situated in the Perseus spiral arm, the system is easily visible on a clear night under low light pollution conditions and represents one of the richest and brightest stellar aggregates in the Milky Way  \citep{capilla02}. Due to its high population of evolved massive stars, including red supergiants, B giants, and early B-type dwarfs, the system is young, with an estimated age below 30 Myr~\citep{currie10}.

Both clusters have been extensively studied using spectroscopic, photometric, and astrometric data.  Binary clusters are generally defined as pairs that share similar 3D positions and space velocities, being gravitationally bound, and frequently assumed to be coeval, i.e., formed in a single star formation event within the same giant molecular cloud \citep{fujimoto97, bekki04}. Due to their apparent spatial proximity and similar physical properties, including age, metallicity, and reddening (\cite{hunt2024, dia21, keller01, capilla02, slesnick2002, Li2019, cantat2020, zhong19, currie10, points04}), NGC 869 and NGC 884 have long been considered the most well-known example of a primordial binary system \cite{messow, slesnick2002, zhong19, liu25}. However, the question of whether they are gravitationally bound or not is still open to discussion. Despite extensive observational and theoretical efforts for more than one century, the true dynamical nature of h and $\chi$ Persei remains debated. For example, \cite{wang24} stated that the two clusters co-evolve from a common origin scenario and are dynamically bound. On the other hand, \cite{uribe02} suggested a possible disruptive pathway from tangential velocity analysis.

The observed astrophysical similarities between the two clusters indeed suggest a common formation, but may not necessarily imply a gravitationally bound binary system. Instead, they may represent two stellar aggregates formed within the same star-forming complex and currently members of a larger expanding association. \cite{dalessandro21} found a primordial group of seven co-moving clusters including the double cluster, called LISCA I. All of them exhibit similar physical properties and the authors stated that the group might be merging. Thus, the resemblance between NGC 869 and NGC 884 could result from co-formation from the same molecular cloud. This system can be considered a paradigmatic example of the Primordial Group hypothesis \citep{juan22}.  

 Thus, while several works argue that the pair constitutes a gravitationally bound binary cluster, others emphasize its origin as part of a larger star-forming complex with dynamical evolution driven by environmental effects.  In this study, we revisit this ongoing debate by investigating the present-day dynamical state and physical relation between NGC 869 and NGC 884, including their structural, kinematic, and stellar population properties. We perform both observationally-based research and $N$-body simulations in order to determine whether they represent a true binary cluster system or a co-formed but unbound pair.

We present new CCD photometry of NGC 869 and NGC 884, which has been combined with photometric and astrometric data from the third data release of the \textit{Gaia} mission (\textit{Gaia} DR3) \citep{GaiaDR3}. Their astrophysical parameters (colour excess, distance, and age) have been determined from CCD UBVRI and \textit{Gaia} DR3 photometric data, and thus evaluated to determine the evolutionary fate of both OCs.

This paper is organised as follows. In Section~\ref{sec:data}, we describe the data and observations used in this study. In Section~\ref{sec:methods}, we present the methodology to determine the member stars and attempt to define the spatial limits of the open clusters. In Section~\ref{sec:results}, we derive the astrophysical parameters obtained within the framework of our analysis. In Section~\ref{sec:dynamics}, we investigate the kinematics and dynamical state of the system and discuss its possible future evolution. Finally, the main conclusions of this study are summarised in Section~\ref{sec:conclusions}.

\section{DATA}
\label{sec:data}

For this study, we used \textit{Gaia} DR3 \citep{GaiaDR3}, complemented by CCD photometry obtained at the Sierra San Pedro Mártir Observatory (Mexico), as well as spectroscopic radial velocities from the LAMOST survey \citep{zhao:2012}, a wide-field, multi-object facility providing medium-resolution optical spectra well suited for studies of open clusters. The combination of \textit{Gaia} astrometric and photometric data with ground-based CCD observations, supplemented by line-of-sight velocity information, allows us to perform a thorough analysis of the cluster membership, astrophysical parameters, and colour--magnitude diagrams (CMDs). Here, we present the CCD photometric observations and their reduction process with a detailed description of the \textit{Gaia} DR3 data retrieval and selection criteria.

\subsection{\textit{Gaia} DR3 and LAMOST} \label{sec:gaia}

For each cluster, \textit{Gaia} DR3 data were extracted from the \textit{Gaia} archive\footnote{https://gea.esac.esa.int/archive/} within a circular region of radius $r = 2^\circ$, centred at $(\alpha,\delta) = (34.765^\circ, 57.140^\circ)$ for NGC~869 and $(35.519^\circ, 57.141^\circ)$ for NGC~884 \citep{hunt2024}. The relatively large extraction region was adopted to ensure adequate coverage of the cluster outskirts and surrounding field, following the wide-field approach of \citep{Angelo2023, guido23}. To ensure reliable astrometric and photometric measurements, we retained only sources with renormalised unit weight error $\mathrm{RUWE} < 1.4$ \citep{lindegren21}, resulting in 114,439 and 113,094 stars for NGC~869 and NGC~884, respectively. To correct for systematic biases in \textit{Gaia} parallaxes, we applied the zero-point correction following \citet{lindegren21} using the \texttt{gaiadr3\_zeropoint} Python package\footnote{https://pypi.org/project/gaiadr3-zeropoint/}.

After selecting members for both clusters following the methods described in Section~\ref{sec:methods}, spectroscopic radial velocities from the LAMOST survey were used to complement the \textit{Gaia} astrometric data by providing line-of-sight velocity information for a subset of cluster members. A more detailed description of the velocity analysis is presented in Section~\ref{sec:dynamics}.

\subsection{CCD Photometry: Observations and Data Reduction}
	
NGC~869 was observed on 2016 February 5 using the 0.84\,m telescope equipped with a Marconi~3 CCD detector (2048$\times$2048 pixels, 13.5\,$\mu$m pixel size) at a binning of 2$\times$2, corresponding to a plate scale of $\sim$0.43\,arcsec\,pixel$^{-1}$. The CCD has a gain of 1.71\,e$^{-}$/ADU and a readout noise of 4.9\,e$^{-}$. NGC~884 was observed earlier, on 2010 September 17, using the 1.5\,m telescope at the same observatory, also equipped with the Marconi~3 CCD camera. The CCD was operated with a binning of 2$\times$2, providing a uniformly illuminated field of view of $\sim$7.9$\times$7.9\,arcmin$^{2}$, a gain of 1.71\,e$^{-}$/ADU, and a readout noise of 4.9\,e$^{-}$. Both clusters were observed through the standard Johnson--Cousins $UBVRI$ filters with a series of short, medium, and long exposures to ensure accurate photometry over a wide magnitude range.
	
Standard star fields \citep{Landolt2009} were also observed near the meridian and at about two airmasses, to determine the atmospheric extinction coefficients and transformation equations. The observing log (Table~\ref{tab:summary}) lists the central equatorial and Galactic coordinates, airmass range, and exposure times in each filter. Bias and flat-field frames were obtained at the beginning and end of each night. Data reduction was carried out using standard procedures within the IRAF/DAOPHOT\footnote{IRAF is distributed by the National Optical Observatories, operated by the Association of Universities for Research in Astronomy, Inc., under cooperative agreement with the National Science Foundation.} package \citep{stet87}. 

\renewcommand{\tabcolsep}{1mm}
\renewcommand{\arraystretch}{1.1}
\begin{table}
    \centering
    \caption{Equatorial/Galactic coordinates and observation summary of NGC 869 and NGC 884.}
    \label{tab:summary}
    \begin{tabular}{lrr}
        \hline
        Parameter & NGC 869 & NGC 884 \\
        \hline
        $\alpha(2000)$ (h m s)    & 02 19 00.0~~$(34^{\circ}.749)$ & 02 22 4.13~~$(35^{\circ}.517)$ \\
        $\delta(2000)$ ($^{\circ}$ $'$ $''$) & +57 08 19.7~~$(+57^{\circ}.139)$ & +57 08 0.6~~$(+57^{\circ}.134)$ \\
        $\ell$ ($^{\circ}$)       & 134.626 & 135.052 \\
        $b$ ($^{\circ}$)          & -3.737 & -3.568 \\
        \hline
        Airmass                   & 1.203 -- 1.282 & 1.114 -- 1.118 \\
        $U$ (Exp. Time s)         & 3, 100, 1000 & 4, 5, 600 \\
        $B$ (Exp. Time s)         & 1, 50, 500 & 2, 20, 150 \\
        $V$ (Exp. Time s)         & 1, 30, 300 & 1, 20, 120 \\
        $R$ (Exp. Time s)         & 0.3, 20, 200 & 0.5, 20, 100 \\
        $I$ (Exp. Time s)         & 0.2, 20, 200 & 0.5, 10, 100 \\
        Cluster stars             &368         & 174 \\
        \hline
    \end{tabular}
\end{table}

Standard magnitude for a given filter $\lambda$ is obtained using the following relation

\begin{equation}
    M_{\lambda} = m_{\lambda} - [k_{1\lambda} -k_{2\lambda}C] X + \eta_{\lambda} C + \zeta_{\lambda}
\end{equation} 
	
where $m_{\lambda}$, $k_{1\lambda}$, $k_{2\lambda}$, $C$, and $X$  are the observed instrumental magnitude, primary/secondary extinction coefficients, colour index and air mass, respectively. $M_{\lambda}$, $\eta_{\lambda}$, $\zeta_{\lambda}$ are standard magnitude, transformation coefficient and photometric zero point, respectively.  More details on data reduction, extinction coefficients and zero points for $UBVRI$ filters can be found in \cite{akk10}, \cite{akk15} and \cite{akk24}. The mean photometric errors in $V$ and colours  $(R-I)$, $(V-I)$, $(B-V)$, $(U-B)$ against  $V$-mag intervals  for NGC~869 and NGC~884 OCs are listed in Table~\ref{tab:pherr} for 368 stars observed in the field of NGC 869, and 174 stars in the field of NGC 884. Stars towards fainter magnitudes have errors less than $\sim0.05$ mag. 

\renewcommand{\tabcolsep}{3mm}
\renewcommand{\arraystretch}{1.0}
\begin{table}
    \centering
    \caption{Mean photometric errors of $V$, $(R-I)$, $(V-I)$, $(B-V)$ and $(U-B)$ for NGC~869 and NGC~884 in terms of $V$ mag.}
    \label{tab:pherr}
    \begin{tabular}{cccccc}
        \hline
        \multicolumn{6}{c}{\textbf{NGC 869}} \\
        \hline
        V &$\sigma_{V}$&$\sigma_{R-I}$&$\sigma_{V-I}$ &$\sigma_{B-V}$ &$\sigma_{U-B}$ \\
        \hline
        6 - 8 & 0.119 & 0.237 & 0.136 & 0.170 & 0.251 \\
        8 - 9 & 0.060 & 0.163 & 0.069 & 0.134 & 0.180 \\
        9 - 10 & 0.066 & 0.132 & 0.072 & 0.124 & 0.162 \\
        10 - 12 & 0.041 & 0.085 & 0.044 & 0.091 & 0.119 \\
        12 - 13 & 0.024 & 0.048 & 0.026 & 0.056 & 0.071 \\
        13 - 14 & 0.017 & 0.040 & 0.020 & 0.041 & 0.045 \\
        14 - 15 & 0.016 & 0.027 & 0.017 & 0.032 & 0.036 \\
        15 - 16 & 0.013 & 0.020 & 0.016 & 0.022 & 0.034 \\
        16 - 17 & 0.013 & 0.016 & 0.015 & 0.022 & 0.034 \\
        17 - 18 & 0.018 & 0.018 & 0.021 & 0.029 & 0.048 \\
        \hline
        \multicolumn{6}{c}{\textbf{NGC 884}} \\
        \hline
        V &$\sigma_{V}$&$\sigma_{R-I}$&$\sigma_{V-I}$ &$\sigma_{B-V}$ &$\sigma_{U-B}$ \\
        \hline
        7 - 9 & 0.030 & 0.064 & 0.031 & 0.031 & 0.011 \\
        9 - 10 & 0.042 & 0.092 & 0.046 & 0.048 & 0.023 \\
        10 - 11 & 0.027 & 0.058 & 0.029 & 0.034 & 0.021 \\
        11 - 12 & 0.023 & 0.047 & 0.026 & 0.029 & 0.020 \\
        12 - 13 & 0.016 & 0.030 & 0.019 & 0.020 & 0.014 \\
        13 - 14 & 0.011 & 0.021 & 0.015 & 0.014 & 0.012 \\
        14 - 15 & 0.008 & 0.015 & 0.011 & 0.011 & 0.013 \\
        15 - 16 & 0.007 & 0.008 & 0.009 & 0.011 & 0.016 \\
        16 - 17 & 0.007 & 0.008 & 0.009 & 0.013 & 0.024 \\
        17 - 18 & 0.012 & 0.013 & 0.015 & 0.023 & 0.035 \\
        \hline
    \end{tabular}
\end{table}
	
\begin{figure*}
 	\centering
 	\includegraphics[width=8.5cm]{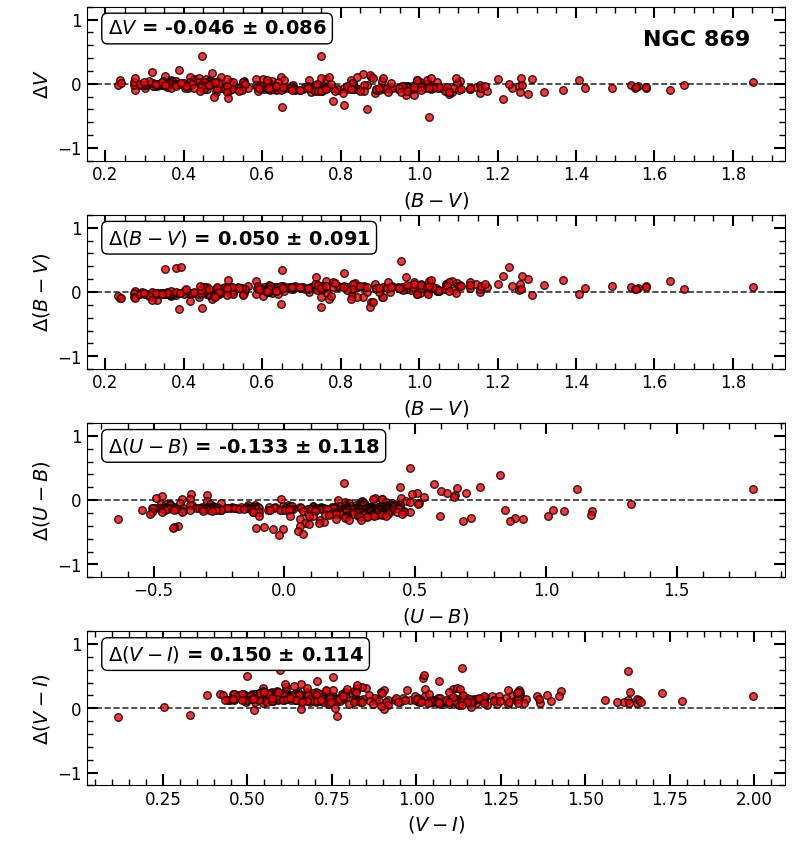}
 	\includegraphics[width=8.5cm]{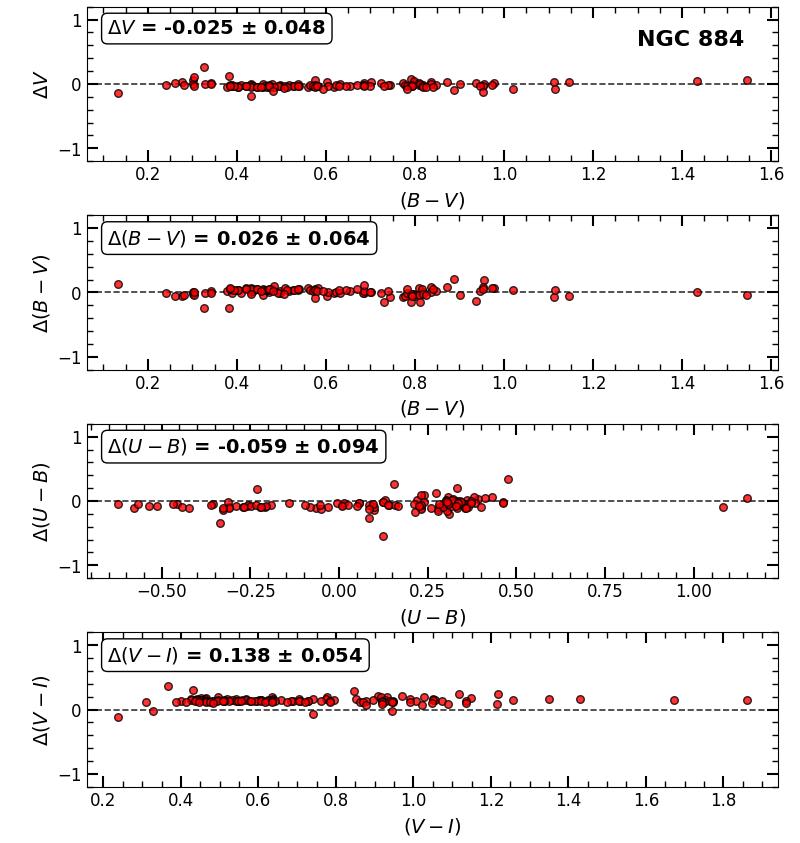}
 	\caption{Photometric differences between our CCD photometry and the measurements of \citet{keller01} for NGC~869 (left) and NGC~884 (right). The differences in $V$, $(B-V)$, $(U-B)$, and $(V-I)$ are plotted as a function of the corresponding colour indices, with $\Delta$ denoting our measurements minus those of \citet{keller01}. The mean offsets and  standard deviations are indicated in each panel. The comparison shows relatively small offsets in $V$ and $(B-V)$, whereas systematic differences are more evident in $(U-B)$ and $(V-I)$.}
 	\label{fig:ubv_kel}
 \end{figure*}

The photometry of NGC 869 and NGC 884 was compared with the $UBVI$ CCD photometry of  \citet{keller01}. To ensure accurate identification, we performed a cross-match between our coordinates and their data, utilising a maximum spatial separation of 1 arcsec. This procedure yielded successful matches for 336 stars in NGC 869 and 124 stars in NGC 884. The resulting differences in $V$, $(B-V)$, $(U-B)$, and $(V-I)$ as a function of colour indices are presented in Figure~\ref{fig:ubv_kel}. For NGC 869 and NGC 884, the mean differences and standard deviations in $V$ and $(B-V)$ are $(\Delta V = -0.046 \pm 0.086, -0.025 \pm 0.048)$ and $(\Delta (B-V) = 0.050 \pm 0.091, 0.026 \pm 0.064)$, respectively. These values indicate that our $V$ magnitudes are slightly brighter and our $(B-V)$ colours are marginally redder than those reported by \citet{keller01}. In the case of $(U-B)$, the mean offsets are $\Delta (U-B) = -0.133 \pm 0.118$ for NGC 869 and $-0.059 \pm 0.094$ for NGC 884, suggesting that our measurements are slightly bluer on average. Conversely, the $(V-I)$ comparisons show mean differences of $\Delta (V-I) = 0.150 \pm 0.114$ and $0.138 \pm 0.054$, respectively, indicating that our colours are systematically redder than those of \citet{keller01}. Thus, while the two datasets show generally consistent photometric behaviour, measurable zero-point offsets are present, particularly in $(U-B)$ and $(V-I)$. No strong colour-dependent trends are evident in the residuals, suggesting that these differences are primarily systematic offsets rather than pronounced colour-dependent effects. The dispersion of the residuals is also notably smaller for NGC~884 than for NGC~869. To assess whether these offsets could affect the derived astrophysical parameters, we explicitly tested their impact on the reddening and CMD-fitting results (see Section~\ref{sec:astro-params}).

The \textit{Gaia} DR3 and $UBVRI$ CCD datasets play complementary roles in this analysis. Ground-based $UBVRI$ CCD photometry covers the central regions and provides crucial leverage for determining interstellar reddening, distance, and age via $(U-B)$ versus $(B-V)$ colour--colour and CMD \citep{bisht19, dia22}. Conversely, \textit{Gaia} DR3 offers wide-field astrometry necessary for reliable member identification, structural modelling, and stellar mass derivation for dynamical analysis. Cross-checking the CCD-derived astrophysical parameters against \textit{Gaia} CMDs yields robust constraints for young, heavily reddened open clusters such as NGC~869 and NGC~884 \citep{karatas23}.

\section{Methods}\label{sec:methods}
	
\subsection{Structural Parameters}\label{sec:struct-params}

The structural parameters were determined from the \textit{Gaia} DR3 stellar distribution prior to the HDBSCAN membership selection. For each cluster, all \textit{Gaia} sources satisfying the astrometric and photometric quality criteria described in Section~\ref{sec:gaia}, comprising 114,439 and 113,094 stars for NGC~869 and NGC~884, respectively, were used to construct the radial density profiles (RDPs). This approach allows the spatial extent of each cluster to be characterised over the full \textit{Gaia} field without being restricted to the smaller area covered by the CCD observations. Stars were binned into concentric annuli of width $0.5\arcmin$. The uncertainties on the stellar surface densities were estimated using Poisson statistics. We fitted the observed RDP using non-linear least-squares optimisation (via \texttt{scipy.optimize.curve\_fit}) with~\citet{king:1962}, since it describes the outer regions of the cluster. It is defined as
	\begin{equation}
		\begin{split}
			\sigma_{R} = 
			\begin{cases}  
				\sigma_{bg} + k \Bigg[ \frac{1}{\sqrt{1+(R/R_c)^2}} 
				- \frac{1}{\sqrt{1+(R_t/R_c)^2}} \Bigg]^{2}, & R \leq R_t, \\[1ex]
				\sigma_{bg}, & R > R_t,
			\end{cases}
		\end{split}
		\label{eq:king62}
	\end{equation}
	where $\sigma_{\mathrm{bg}}$ is the background stellar density, $R_c$ is the core radius (the radius at which the density falls to half the central value), and $R_t$ is the tidal radius—interpreted as the distance where the cluster’s gravitational influence is balanced by external tidal forces, i.e., where $\sigma(R_t) = \sigma_{\mathrm{bg}}$. The limiting radius ($R_{\mathrm{lim}}$), taken as the observational truncation radius, was derived from the best-fit parameters and characterises the spatial extent of the cluster above the field background. For uncertainties on structural parameters, we initially estimated formal uncertainties from the covariance matrix returned by the non-linear least-squares fitter (\texttt{scipy.optimize.curve\_fit}), as $\sigma_{\theta_j} = \sqrt{\mathrm{Cov}_{jj}}$, where $\theta_j$ denotes the $j$-th fitted parameter. However, to account for potential non-Gaussianity and heteroscedasticity in the residuals, we also implemented a bootstrap resampling approach, which provides confidence intervals that are less sensitive to assumptions about error normality and weighting.

	\subsection{Membership Selection}\label{sec:membership-selec}

HDBSCAN is an enhanced version of the DBSCAN algorithm \citep{est1996, camp2013, mci2017}. Unlike DBSCAN, which requires specifying a global distance threshold and the minimum number of points, HDBSCAN only requires two parameters: the minimum cluster size $m_{clSize}$, and the minimum number of points in the neighbourhood of a cluster core point $m_{Pts}$. HDBSCAN has previously been applied to identify moving groups in \textit{Gaia} data by \cite{koun2019} and \cite{koun2020}. 

\begin{figure}
	\centering{\includegraphics[width=0.92\columnwidth]{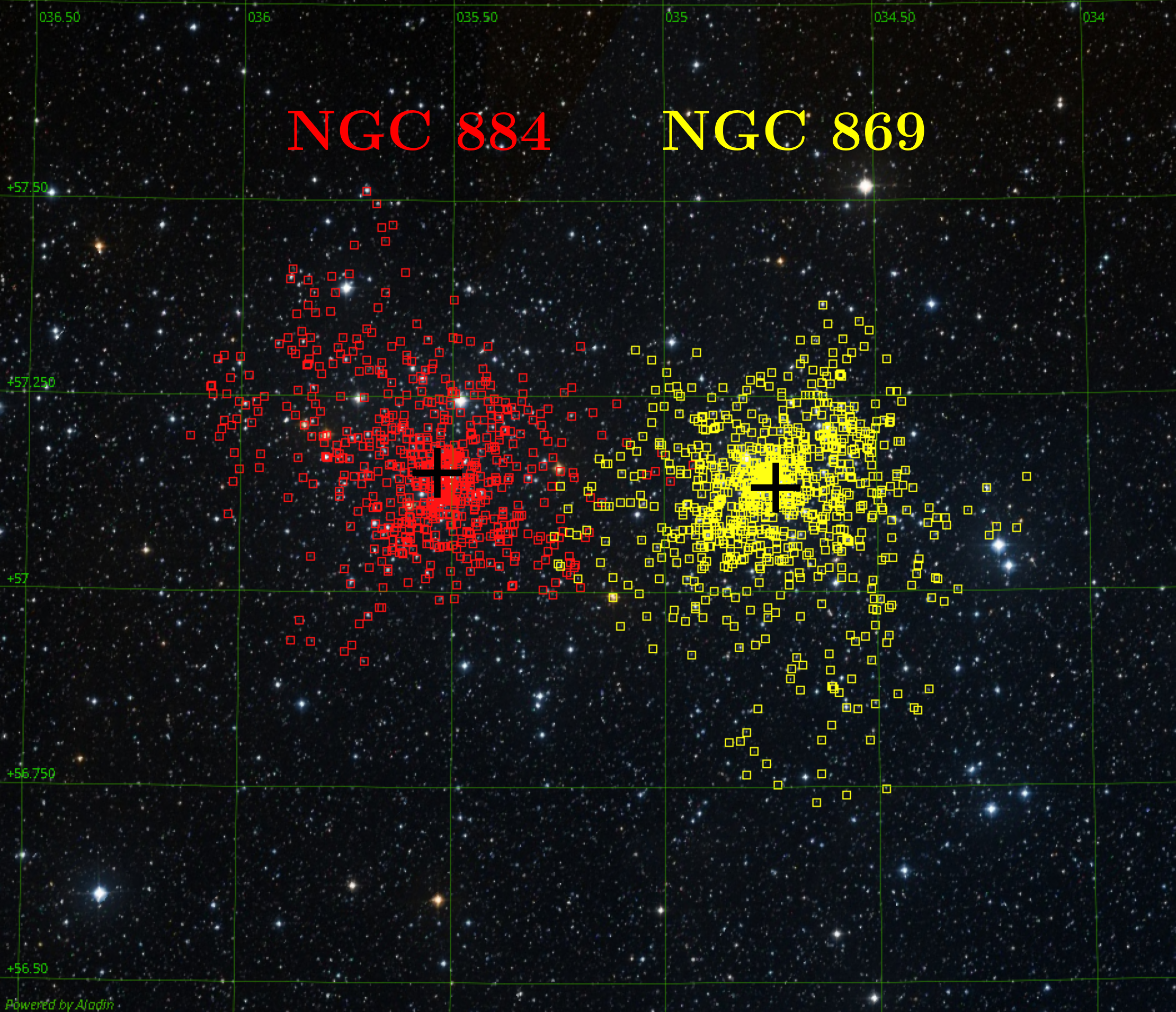}} \vspace{-2mm}
		\caption{The RA, DEC skymap position ($3^{\circ}$ × $1.3^{\circ}$) of the NGC 869 and NGC 884 pair from the SIMBAD-Aladin (\url{https://aladin.cds.unistra.fr/aladin.gml}). North is up, and east is left. Yellow and red squares represent the likely members of NGC 869 and NGC 884, respectively.}
	\label{figure_3}
\end{figure}

We used the HDBSCAN clustering algorithm 
with the leaf cluster-selection
method in five dimensions ($\ell$, $b$, $\mu_\alpha$, $\mu_\delta$, and $\varpi$), adopting $m_{clSize}=40$ and $m_{Pts}=10$. We obtained 830 and 644 probable members of NGC 869 and NGC 884, respectively, with a membership probability greater than 0.75 for both clusters (see Figure~\ref{figure_3}). Previous studies (e.g., \citealt{zhong19}; \citealt{currie10}) suggest that members of NGC 869 and NGC 884 may extend beyond the clusters’ limiting radius $R_{lim}$ and tidal radius $R_{t}$. Therefore, additional bona fide members may exist outside $R_{lim}$ that are not included in our current membership sample.

Because the resulting membership sample is subsequently used for the kinematic and dynamical analyses, we tested the sensitivity of the detected cluster populations to the adopted $m_{clSize}$, while keeping $m_{Pts}=10$ fixed. We considered $m_{clSize}=10, 20, 30, 40, 50, 60, 70, 80,$ and 100. The values $m_{clSize}=10, 20, 40,$ and 80 are also within the range adopted by \cite{hunt23} in their Gaia-based open-cluster search. For $m_{clSize}=10$ and 20, HDBSCAN separates the cluster populations into several smaller groups, indicating substantial fragmentation. At $m_{clSize}=30$, we recover 729 and 578 probable members for NGC 869 and NGC 884, respectively. For $m_{clSize}=40, 50, 60, 70, 80,$ and 100, however, HDBSCAN recovers exactly the same member lists as those adopted in our original analysis, namely 830 members for NGC 869 and 644 members for NGC 884.

This behaviour indicates that the adopted $m_{clSize}=40$ lies within a stable regime of the HDBSCAN solution. In particular, the identical membership lists obtained over the range $m_{clSize}=40--100$ demonstrate that the adopted membership sample is not sensitive to moderate-to-large changes in the minimum cluster size. The fragmentation observed for $m_{clSize}=10$ and 20, followed by the stable solution for $m_{clSize}\geq40$, supports the identification of the two main cluster populations and provides confidence that the subsequent kinematic and dynamical results are not driven by the particular choice of $m_{clSize}=40$.

\section{Results}
\label{sec:results}

\subsection{Membership and Structural Properties}

\begin{figure*}
    \centering
    \includegraphics[width=0.45\textwidth]{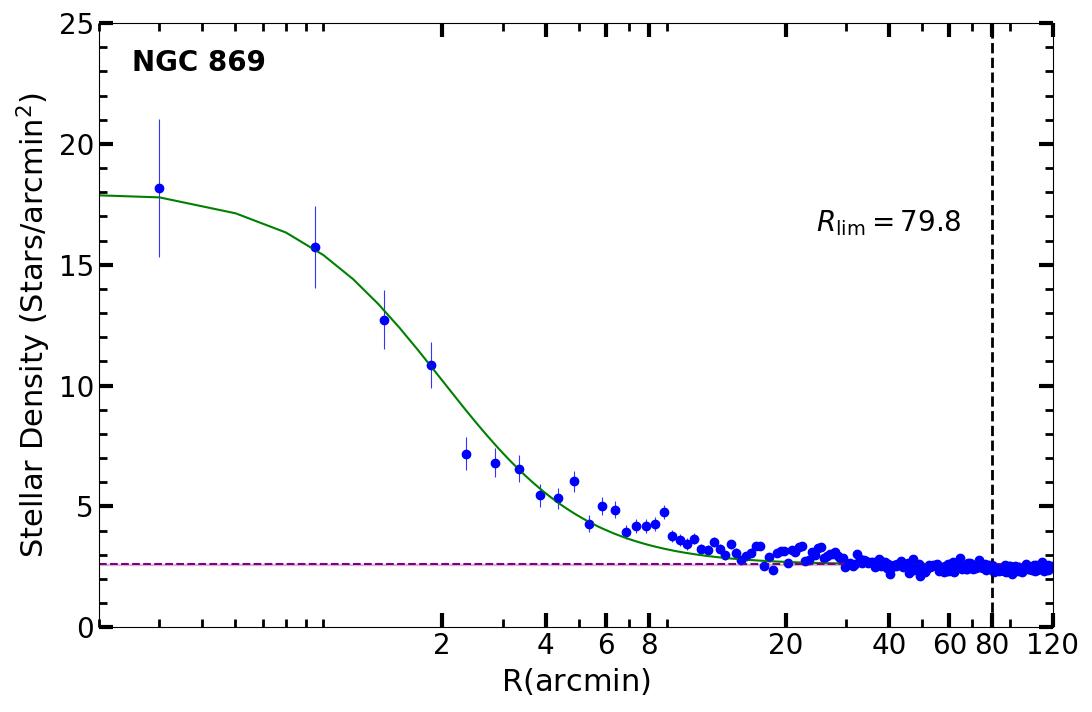}
    \includegraphics[width=0.45\textwidth]{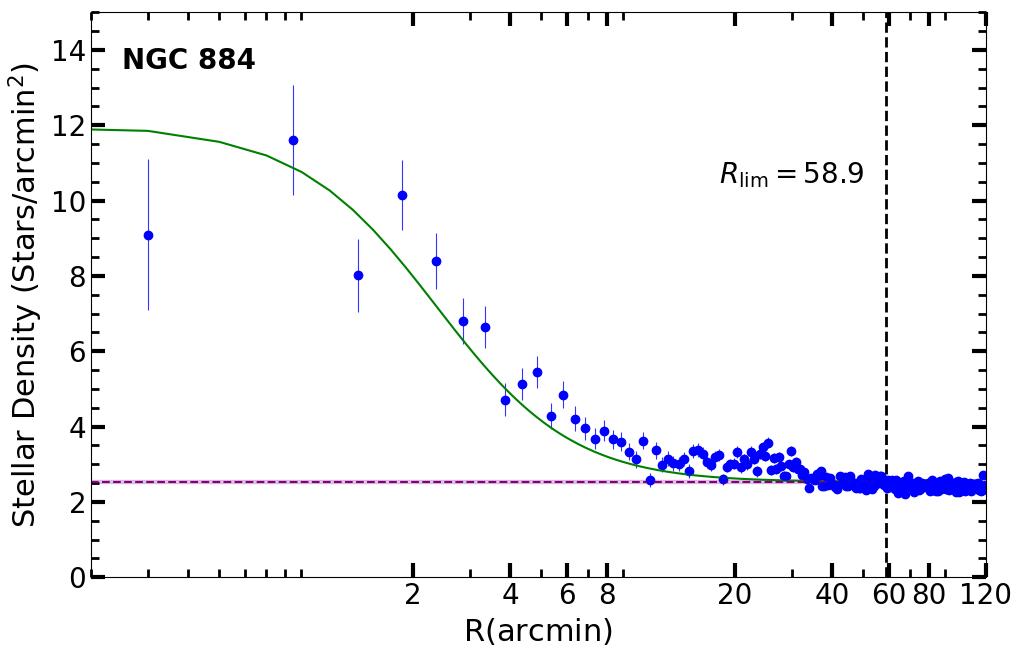}
    \caption{Radial density profiles of NGC~869 (left panel) and NGC~884 (right panel), constructed from the \textit{Gaia} DR3 sources within the $2^\circ$-radius extraction fields. The solid curves show the best-fit King profiles, while the horizontal dashed lines indicate the stellar background levels measured in the comparison fields. The vertical dashed lines mark the derived limiting radii, $R_{\rm lim}$. The error bars represent the Poisson uncertainties, estimated as $\sqrt{N}$ for each annulus. The wide radial coverage allows the outer stellar distributions to be traced well beyond the cluster cores and provides sufficient coverage for determining the limiting radii.}
    \label{fig:RDP}
\end{figure*}

The resulting radial density profiles (RDP) for both clusters are given in Figure \ref{fig:RDP}. The limiting radii were determined to be 80$\pm$1 arcmin and 59$\pm$2 arcmin for NGC~869 and NGC~884, respectively, while the tidal radii were found to be 102$\pm$6 arcmin and 82$\pm$7 arcmin, respectively. The central stellar densities are $16.08\pm0.63$ and
$10.00\pm0.63$ stars arcmin$^{-2}$ for NGC~869 and NGC~884,
respectively, while the corresponding background densities are
$2.60\pm0.03$ and $2.53\pm0.03$ stars arcmin$^{-2}$. The derived structural parameters of NGC~869 and NGC~884, expressed in both arcmin and parsecs, along with literature values, are listed in Table~\ref{tab:tbl-struct}. The conversion from arcmin to parsecs was performed using parallaxes obtained for the clusters, as given in Table~\ref{tab-PM}.

To assess whether the relatively large structural radii are sensitive to the adopted radial coverage, we repeated the RDP analysis using field radii of $90\arcmin$ and $120\arcmin$ around each cluster. The resulting limiting and tidal radii are consistent between the two field sizes for both NGC~869 and NGC~884, indicating that the derived structural parameters are not significantly affected by the adopted outer radial coverage. These results are also consistent with the extended structures reported for the Double Cluster by \cite{zhong19}. We therefore adopted the $120\arcmin$ field for the final structural analysis to ensure sufficient coverage of the outer cluster profiles.

	\renewcommand{\tabcolsep}{1.5mm}
	\renewcommand{\arraystretch}{1.1}
	\begin{table*}
		\caption{Structural parameters of NGC 869 and NGC 884 together with the literature values. The arcmin to pc conversions are 0.72 (NGC 869) and 0.73 (NGC 884), respectively.}
		\resizebox{1.0\textwidth}{!}{ 
			\centering
			\begin{tabular}{ccccc|cccccc}
				\hline
				\multicolumn{10}{c}{\textbf{NGC 869}} &  \\
				\hline
				{$\sigma_{0}$} & {$\sigma_{bg}$} & {$R_{c}$} & {$R_{lim}$} & {$R_{t}$} & {$\sigma_{0}$} & {$\sigma_{bg}$} & {$R_{c}$} & {$R_{lim}$} & {$R_{t}$} & Ref \\
				{($*\,\prime^{-2}$)} & {($*\,\prime^{-2}$)} & {($\prime$)} & {($\prime$)}&{($\prime$)} &{($pc^{-2}$)} & {($pc^{-2}$)} & {($pc$)} & {($pc$)} & {($pc$)} & \\
				\hline

				$16.08\pm0.63$ & $2.60\pm0.03$ & $2.00\pm0.08$ & $80\pm1$ &$102\pm6$& $31.02\pm2.28$ & $5.02\pm0.06$ & $1.44\pm0.06$ & $58\pm1$ &$73\pm4$& This study \\
				&  & $8.08$ & & $11.94$&  & & $5.36$ & & $7.94$ &1 \\
				&  &  &  & $44.79\pm3.43$ & &  & &  &$32.25\pm2.47$& 2 \\
				& &  & $14.40$ & &  &  & & $10.37$ & & 3 \\
				& &  & $20.70$ & &  &  & & $14.90$ & & 4 \\
				\hline
				\multicolumn{10}{c}{\textbf{NGC 884}} &  \\
				\hline
				{$\sigma_{0}$} & {$\sigma_{bg}$} & {$R_{c}$} & {$R_{lim}$} & {$R_{t}$} & {$\sigma_{0}$} & {$\sigma_{bg}$} & {$R_{c}$} & {$R_{lim}$} & {$R_{t}$} & Ref \\
				{($*\,\prime^{-2}$)} & {($*\,\prime^{-2}$)} & {($\prime$)} & {($\prime$)}&{($\prime$)} &{($pc^{-2}$)} & {($pc^{-2}$)} & {($pc$)} & {($pc$)} & {($pc$)} & \\
				\hline

				$10.00\pm0.63$ & $2.53\pm0.03$ & $2.38\pm0.14$ & $59\pm2$ &$82\pm7$& $18.77\pm1.18$ & $4.75\pm0.06$ & $1.74\pm0.10$ & $43\pm1$ &$60\pm5$& This study \\
				&  & $7.25$ &  & $9.05$&  & & $4.84$ & $6.03$ & $6.61$ &1 \\
				&  &  &  & $54.00\pm11.86$ & &  & &  &$39.42\pm8.66$& 2 \\
				& &  & $10.50$ & &  &  & & $7.67$ & & 3 \\
				& &  & $13.90$ & &  &  & & $10.15$ & & 4 \\
				\hline
			\end{tabular}
		}
		\begin{tabular}{p{18cm}}
			1: \citep{hunt2024}, 2: \citep{just23}, 3: \citep{granada18}, 4: \citep{yu18}
		\end{tabular}
		\label{tab:tbl-struct}
	\end{table*}

\begin{figure*}
	\includegraphics[width=13.5cm]{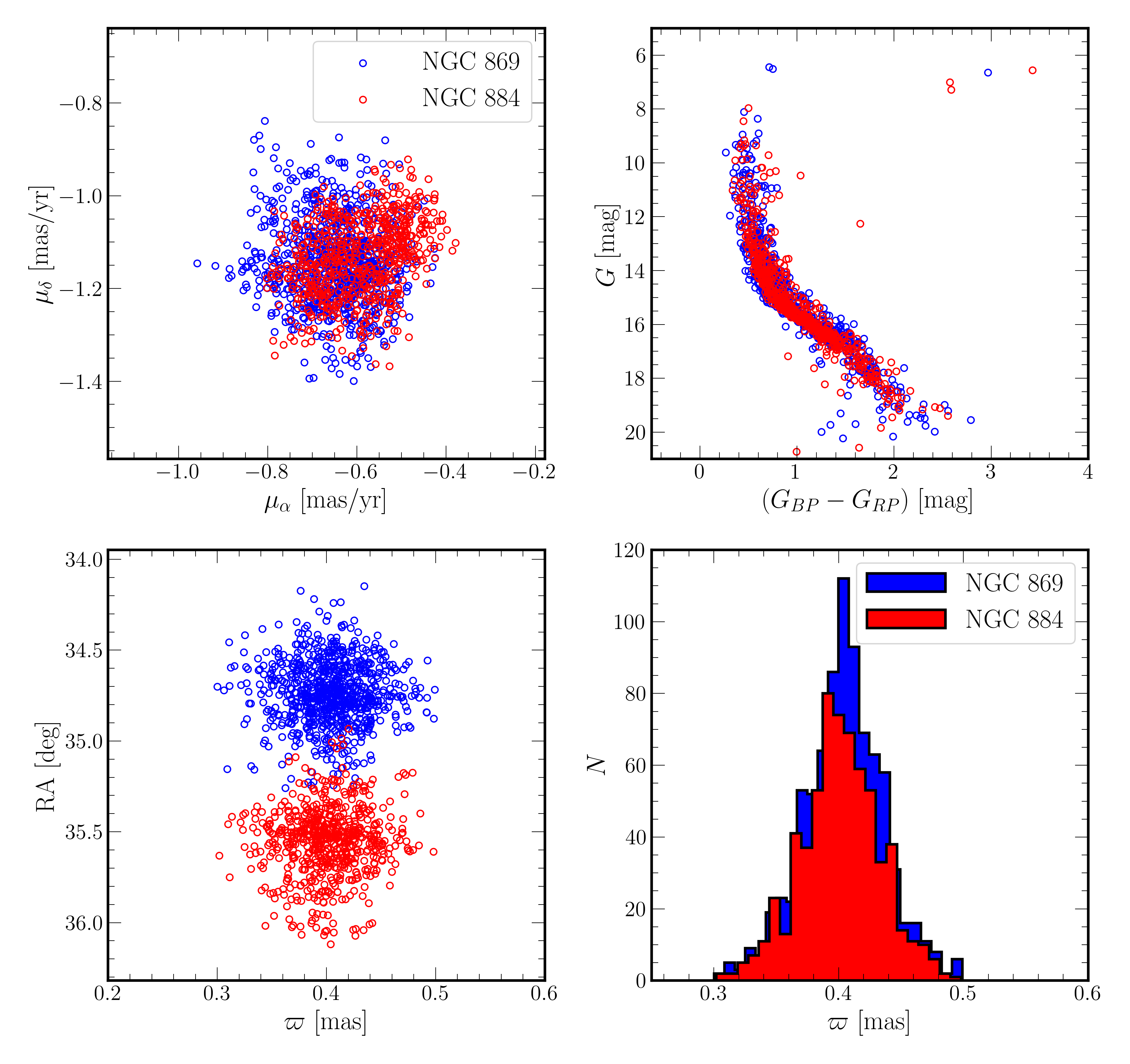}
	\caption{Astrometric and photometric distributions of the probable members identified with HDBSCAN and having membership probabilities $P>0.75$. The upper panels show the joint vector-point diagram (left) and the \textit{Gaia} DR3 CMD (right). The lower panels show $\varpi$ versus $\alpha$ (left) and the corresponding parallax distributions (right). NGC~869 and NGC~884 members are shown in blue and red, respectively.}
	\label{figure_5}
\end{figure*}

The median values of $(\alpha,\delta)$ and  $(\ell, b)$ for the likely members of both OCs are $(34.74^{\circ}$, $57.13^{\circ})$  and $(134.63^{\circ}, -3.74^{\circ})$ for NGC 869 and $(35.54^{\circ}, 57.15^{\circ})$ and  $(135.03^{\circ},-3.58^{\circ})$ for NGC 884, respectively. The vector-point diagram (VPD), the CMDs, the parallax-RA relation, and the parallax distributions of the NGC 869 and NGC 884 are presented in Figure~\ref{figure_5}. The median values of $\mu_{\alpha}$, $\mu_{\delta}$, distance, and $\varpi$ for both clusters are listed in Table~\ref{tab-PM}, together with a comparison to values from the literature. The PMs are slightly but significantly different for both clusters. The parallax of NGC~869 is also slightly higher than that of NGC~884, however, the astrometric distances of the two clusters are mutually consistent.

	The SPM survey of NGC~869 and NGC 884 contains CCD~UBV(RI)$_{KC}$ photometric data of 368 and 174 stars, respectively. Within $R_{lim}$ of both OCs, we cross-matched their coordinates ($\alpha$, $\delta$) and \textit{Gaia} DR3 astrometric/photometric data of the probable members of both OCs. We set 1 arcsec as the maximum difference in ($\alpha$, $\delta$) for merging both datasets. The matching procedure yielded 198 (NGC 869) and 104 (NGC 884) probable members with CCD UBV(RI)$_{KC}$, which are considered for the analyses.

	\renewcommand{\tabcolsep}{1.5mm}
	\renewcommand{\arraystretch}{1.1}
	\begin{table*}
		\caption{Median proper motion components, distance, and parallax of NGC\,869 and NGC\,884. The distances listed in the table are calculated as $d=1/\varpi$, where $\varpi$ is the \textit{Gaia} DR3 parallax given in columns 4 and 8. The literature references are listed in the last column.} 
		\centering
    	\label{tab-PM}
		\begin{tabular}{cccc|ccccc}
			\hline
			\multicolumn{4}{c}{\textbf{NGC 869}} & \multicolumn{5}{c}{\textbf{NGC 884}} \\
			\hline
			$\mu_{\alpha}$ & $\mu_{\delta}$ & $d$ & $\varpi$ & $\mu_{\alpha}$ & $\mu_{\delta}$ & $d$ & $\varpi$ & Ref \\
			(mas/yr) & (mas/yr) & (pc) & (mas) & (mas/yr) & (mas/yr) & (pc) & (mas) & \\
			\hline
			\noalign{\smallskip}
			$-0.642\pm0.004$ & $-1.148\pm0.004$ & $2466\pm6$ & $0.405\pm0.001$ & $-0.590\pm0.007$ & $-1.129\pm0.004$ & $2497\pm10$ & $0.400\pm0.002$ & This study \\
			$-0.654\pm0.078$ & $-1.143\pm0.079$ & $2475\pm116$ & $0.404\pm0.019$ & $-0.613\pm0.079$ & $-1.142\pm0.065$ & $2293\pm118$ & $0.402\pm0.019$ & 1 \\
			$-0.700\pm0.001$ & $-1.100\pm0.001$ & $2500\pm25$ & $0.400\pm0.004$ & $-0.600\pm0.001$ & $-1.100\pm0.001$ & $2500\pm25$ & $0.400\pm0.004$ & 2 \\
			$-0.682\pm0.1192$ & $-1.077\pm0.133$ & $2246\pm64$ & $0.400\pm0.043$ & $-0.626\pm0.125$ & $-1.059\pm0.122$ & $2150\pm103$ & $0.398\pm0.039$ & 3 \\
			$-0.640$ & $-1.170$ & ${2353}_{-197}^{+178}$ & $0.385$ & $-0.640$ & $-1.170$ & ${2353}_{-197}^{+178}$ & $0.385$ & 4 \\
			$-0.653\pm0.100$ & $-1.146\pm0.089$ & $2197$ & $0.407	\pm0.025$ & $-0.612\pm0.113$ & $-1.127\pm0.094$ & $2130$ & $0.402\pm0.025$ & 5\\
			$-0.689\pm0.122$ & $-1.082\pm0.122$ & $2506\pm264$ & $0.399\pm0.042$ & $-0.628\pm0.117$ & $-1.066\pm0.118$ & $2513\pm240$ & $0.398\pm0.038$ & 6 \\
			$ -0.685\pm0.131$ & $-1.074\pm0.146$ & $2344_{-55}^{+88}$ & $0.400\pm0.060$ & $-0.614\pm0.133$ & $-1.058\pm0.134$ & $2344_{-82}^{+87}$ & $0.400\pm0.060$ & 7 \\
			$-0.689\pm0.122$ & $-1.082\pm0.122$ & $2336$ & $0.430$ & $-0.628\pm0.117$ & $-1.066\pm0.118$ & $2342$ & $0.430$ & 8 \\
			
			\hline
		\end{tabular}
		\begin{tabular}{p{18cm}}
			1: \citep{hunt23}, 2: \citep{Song2022}, 3: \citep{dia21}, 4: \citep{Dalessandro2021}, 5: \citep{Poggio2021}, 6: \citep{cantat2020}, 7: \citep{yu23}, 8: \citep{zhong20}. \end{tabular}
		
	\end{table*}

\subsection{ASTROPHYSICAL PARAMETERS OF NGC~869 AND NGC~884}\label{sec:astro-params}
	
Astrophysical parameters were derived for cluster members with CCD~UBVRI photometry obtained at the National Astronomical Observatory San Pedro Mártir. These members were identified by cross-matching two datasets: (1) stars classified as members using the HDBSCAN method (see Section~\ref{sec:membership-selec}), and (2) stars with CCD~UBVRI observations located in the vicinity of each cluster centre. The matching was performed using the $\alpha$, $\delta$, $G_{BP}$, $G_{RP}$, and $G$ magnitudes. As a result, 198 and 104 stars with CCD~UBVRI photometry were confirmed as members of NGC~869 and NGC~884, respectively. The smaller number of members compared to the selection based on \textit{Gaia} DR3 is mainly due to the more limited field of view of the SPM observations. In this section, we determine the reddening, metallicity, age, and distance of these confirmed members.

\subsubsection{Reddening}\label{sec:reddening}

\begin{figure*}
    \begin{center}
    \includegraphics[width=0.38\linewidth]{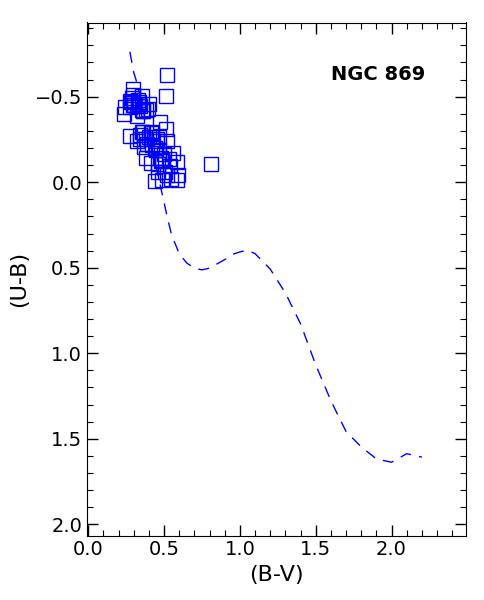} 
    \includegraphics[width=0.38\linewidth]{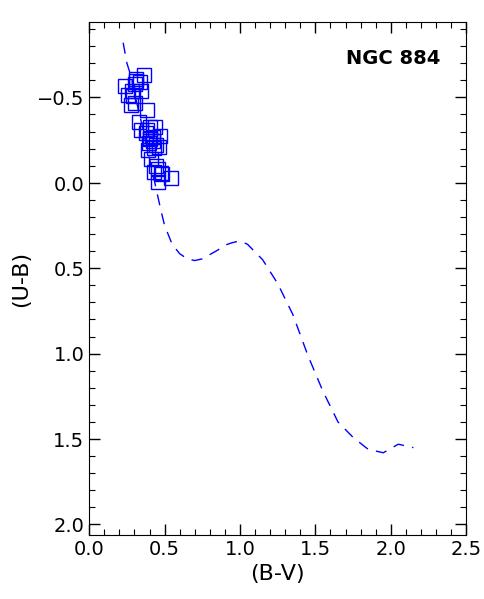}
    \caption{The CC diagrams of NGC~869 (left panel) and NGC~884 (right panel). Open squares represent the early-type members $(U-B) < 0$ (48 in NGC~869 and 37 in NGC~884). The dashed blue lines represent the reddened S13 dwarf sequences.}
        \label{fig:4-CCVUB}
    \end{center}
\end{figure*}

For the reddening determination of NGC~869 and NGC~884, the criterion $\sigma_{(U-B)} < 0.05$ (see Table~\ref{tab:pherr}) has been applied, resulting in 174 members for NGC~869 and 104 members for NGC~884. Early-type members with $(U-B) < 0.0$ (48 for NGC~869 and 37 for NGC~884), represented by blue squares in Figure~\ref{fig:4-CCVUB}, have been used to derive the reddening values of both OCs using the intrinsic colour relations of \cite{sung13}. For NGC~869, the early-type members span the ranges $-0.624 < (U-B) < -0.005$ and $0.240 < (B-V) < 1.230$, while for NGC~884 the corresponding ranges are $-0.755 < (U-B) < -0.007$ and $0.240 < (B-V) < 1.055$. The Q values fall within $(-1.005, -0.326)$ for NGC~869 and $(-1.542, -0.339)$ for NGC~884. The $E(B-V)$ values have been calculated using the relations $(B-V)_0 = 0.328Q - 0.019$, $Q = (U-B) - 0.72(B-V)$, and $E(U-B) = 0.72E(B-V) + 0.025E(B-V)^2$ given by \cite{sung13}. The median reddenings are derived as $E(B-V) = 0.62 \pm 0.10$ mag for NGC~869 and $E(B-V) = 0.60 \pm 0.08$ mag for NGC~884. The reddened S13 colour sequence (blue dashed line) is fitted to the cluster members in Figure~\ref{fig:4-CCVUB}. The selected early-type members of NGC~884 exhibit a relatively tighter distribution around the main sequence compared with NGC~869, where a few stars are slightly displaced
from the \citet{sung13} sequence. 

\begin{figure*}
	\begin{center}
		\includegraphics[width=0.6\linewidth]{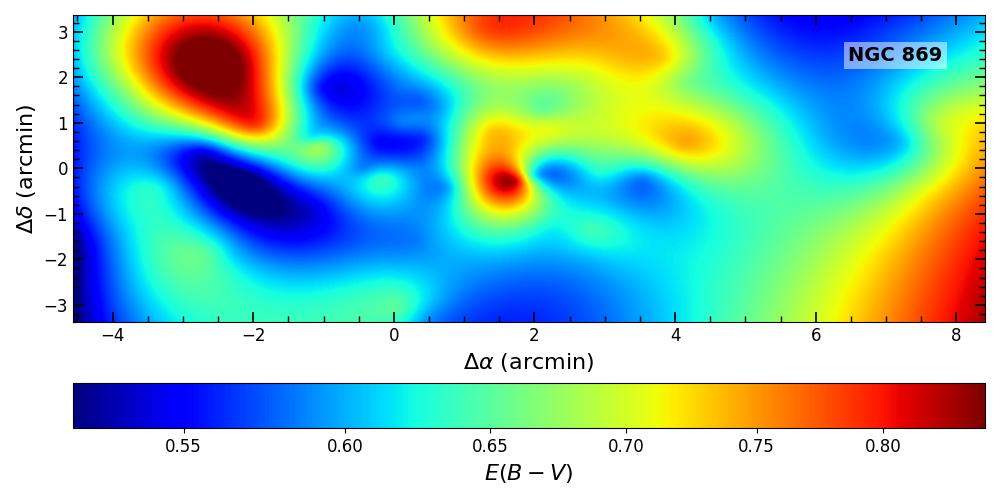} 
		\includegraphics[width=0.6\linewidth]{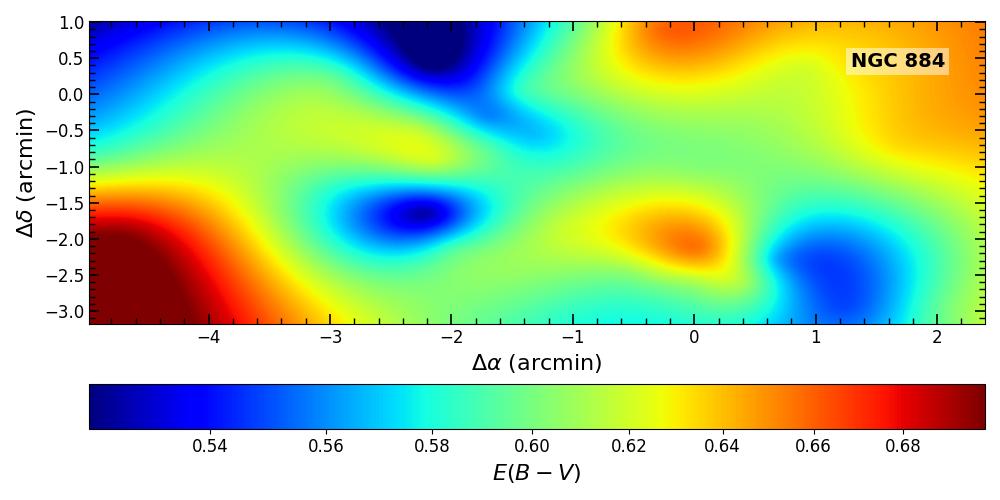}
		\caption{Spatial distribution of the individual $E(B-V)$ values derived for the early-type probable members of NGC~869 (lower panel) and NGC~884 (upper panel) using the $Q$-method. The colour scale represents the individual reddening estimates as a function of the relative right ascension and declination of the early-type members. The maps reveal spatial variations in the reddening across both cluster fields, with a broader range of $E(B-V)$ values in NGC~869 than in NGC~884.}
		\label{fig:dif_red}
	\end{center}
\end{figure*}

The individual reddening estimates show standard deviations of 0.089 and 0.041 mag for NGC 869 and NGC 884, respectively, with NGC 869 exhibiting the larger star-to-star variation. The spatial distributions of these estimates reveal measurable variations across the cluster fields. For NGC 869, the $E(B-V)$ values range from approximately 0.49 to 0.95 mag, whereas those for NGC 884 range from approximately 0.50 to 0.70 mag (see Figure~\ref{fig:dif_red}). These spatial variations are consistent with previous studies reporting differential reddening across the Double Cluster field \citep{keller01,capilla02}. The spatial distributions shown in Figure~\ref{fig:dif_red} are derived from the early-type probable members, which are used as reddening tracers. Since the dispersion of the individual reddening estimates alone does not fully characterise the spatial structure of the extinction, we further investigated its effect on the CMDs (Section~\ref{sec:age-dist}). Bootstrap resampling of the early-type members gives 68\% confidence intervals of 0.60--0.63 mag and 0.60--0.60 mag, and 95\% confidence intervals of 0.60--0.64 mag and 0.60--0.62 mag for the median reddening of NGC 869 and NGC 884, respectively. These narrow intervals indicate that the median reddening estimates are stable against resampling of the early-type members.

\subsubsection{Metallicity, Age and Photometric Distance}\label{sec:age-dist}

Some studies adopted a solar metallicity for NGC~869 and NGC~884 \citep{Li2019, zhong19, Song2022}, while spectroscopic studies by \cite{marsakov16}, \cite{zhong20}, and \cite{Dalessandro2021} gave metal abundances  $[\mathrm{Fe/H}]=-0.30\pm0.10$, $-0.25\pm0.18$, and $-0.29\pm0.03$, respectively. Given the consistency among the available spectroscopic measurements and the higher reliability generally associated with high-resolution spectroscopic metallicities \citep{heiter14}, we adopt $[\mathrm{Fe/H}] = -0.30$ from \cite{marsakov16}. The sensitivity of the derived parameters to the adopted metallicity is examined later in this section. To convert the adopted $[\mathrm{Fe/H}]$ to the metal mass fraction,
we applied
\begin{equation}
	Z = Z_\odot \cdot 10^{[\mathrm{Fe/H}]}
\end{equation}
assuming a solar metal mass fraction of $Z_\odot = 0.0152$ \citep{caffau11}. This results in $Z = 0.008\pm0.002$.

The colour excesses in the photometric bands $E(R-I)$, $E(V-I)$, $E(U-B)$ and $E(G_{\rm BP} - G_{\rm RP})$ were derived from the adopted $E(B-V)$ values using the relations of \cite{dean1978}, \cite{straizys1995}, \cite{sung13}, and \citep{brag2018}, respectively. We adopted
\begin{align}
	E(V-I) &= 1.25\,E(B-V), \\
	E(R-I) &= 0.69\,E(B-V), \\
	E(B-V) &= 0.775\,E(G_{\rm BP} - G_{\rm RP}), \\
	E(U-B) &= 0.72\,E(B-V) + 0.025\,[E(B-V)]^{2}.
\end{align}
The visual extinction was calculated as $A_V=3.1E(B-V)$.

\begin{figure}
	\centering
	\includegraphics[width=0.53\textwidth, trim=0 4mm 0 16mm,
        clip]{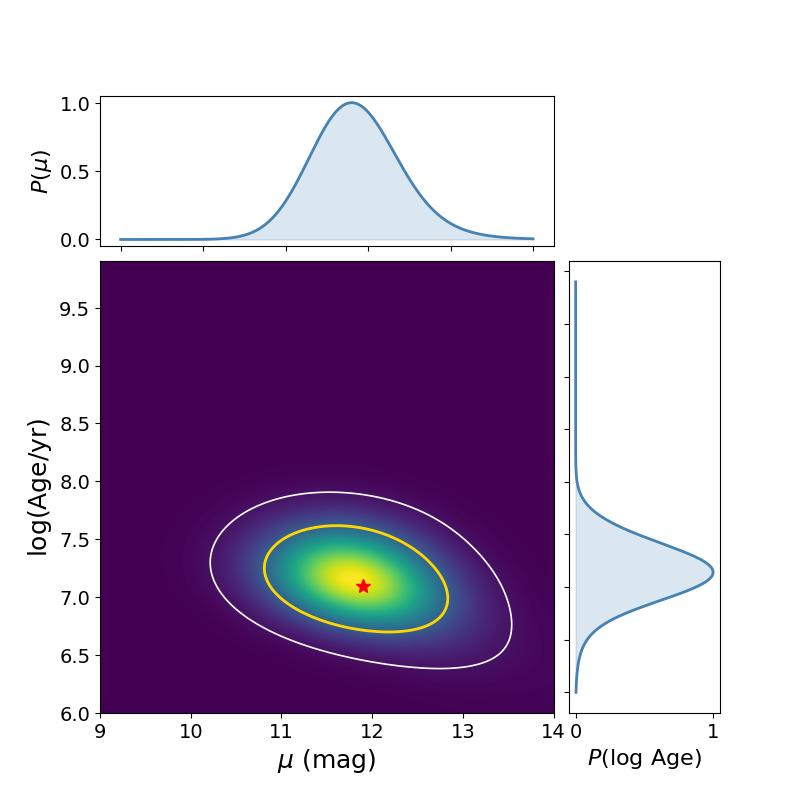}\vspace{-2mm}
    \caption{Joint and marginalised posterior probability distributions for the age, $\log(t/{\rm yr})$, and distance modulus, $\mu$, of NGC~869 obtained from the Bayesian isochrone fitting. The central panel shows the joint posterior distribution, with the red star indicating the maximum-a-posteriori (MAP) solution. The gold and white contours enclose the 68\% and 95\% credible regions, respectively. The top and right panels show the marginalised posterior distributions for age and distance modulus, respectively. The fit assumes $Z=0.008$ and $E(B-V)=0.62$. The corresponding posterior distributions for NGC~884 are essentially indistinguishable and are therefore not shown.}
	\label{fig:bayesian}
\end{figure}

Following the approach of \cite{al2026}, we constrained the ages and distances of NGC~869 and NGC~884 using a Bayesian grid search based on the isochrone-fitting formalism of \citet{Naylor2006}. The fit was performed using 198 and 104 CCD $UBV(RI)_{KC}$ members of NGC~869 and NGC~884, respectively. We explored ages over the range $\log(t/{\rm yr})=6.0$--$10.0$ and true distance moduli, $V_0-M_V$, from 9.0 to 14.0 mag, with a grid spacing of 0.02 in both parameters. For each grid point, the likelihood was evaluated from the $\chi^2$ comparison across the $V$ versus $(B-V)$, $(U-B)$, $(R-I)$, and $(V-I)$ CMDs, with all four CMDs fitted simultaneously for the same age and distance modulus. The resulting posterior distributions were then marginalised to derive the 68\% and 95\% credible intervals. We used version 1.2S of the \texttt{CMD} tool\footnote{\url{https://stev.oapd.inaf.it/cgi-bin}} together with non-rotating PARSEC isochrones from \citet{bressan2012}, adopting a fixed metallicity of $Z=0.008$ and reddening values of $E(B-V)=0.62$ and $0.60$ mag for NGC~869 and NGC~884, respectively. We note that these reddening values were held fixed during the Bayesian grid search and their uncertainties were not directly propagated into the posteriors. To assess this systematic effect, we repeated the fitting procedure by varying $E(B-V)$ by its estimated uncertainty, finding that shifts in reddening introduce systematic variations of $\sim 0.1$ in $\log(t/\text{yr})$ and $0.1\text{--}0.3$ mag in the distance modulus. The maximum-a-posteriori solutions yield $\log(t/{\rm yr})=7.08\pm0.02$ and $7.08\pm0.03$ for NGC~869 and NGC~884, respectively, corresponding to ages of $12\pm1$ Myr for both clusters. The distance moduli are $11.80\pm0.22$ and $11.82\pm0.21$ mag, yielding distances of approximately $2.29\pm0.23$ and $2.29\pm0.22$ kpc. Since the posterior distributions for the two clusters are nearly identical, we show only the posterior distributions for NGC~869 in Figure~\ref{fig:bayesian}.

As an independent check, we tested whether the astrophysical parameters derived from the CCD $UBV(RI)_{KC}$ photometry are consistent with the \textit{Gaia} data. We performed a simultaneous $\chi^2$ optimisation for the 830 and 644 Gaia-selected members of NGC~869 and NGC~884, respectively, using the \texttt{simul\_fit} code\footnote{\url{https://github.com/speletry/simul_fit.py}}. The fit was performed using the $G$ versus $(G_{\rm BP}-G_{\rm RP})$ CMD, with age and distance modulus optimised simultaneously, while the reddening and metallicity were fixed to the adopted values. For NGC~869, the fit yields $\log(t/{\rm yr})=7.10\pm0.01$ and $V_0-M_V=11.70\pm0.05$ mag, while for NGC~884 it gives $\log(t/{\rm yr})=7.10\pm0.01$ and $(V_0-M_V)=11.85\pm0.05$ mag. The Gaia-based age and distance estimates are consistent with those derived from the CCD photometry, supporting the applicability of the adopted reddening and metallicity values to the \textit{Gaia} CMDs. We therefore adopt the CCD-based astrophysical parameters as our primary estimates.

\begin{figure*}
	\centering
	\includegraphics[width=0.265\textwidth]{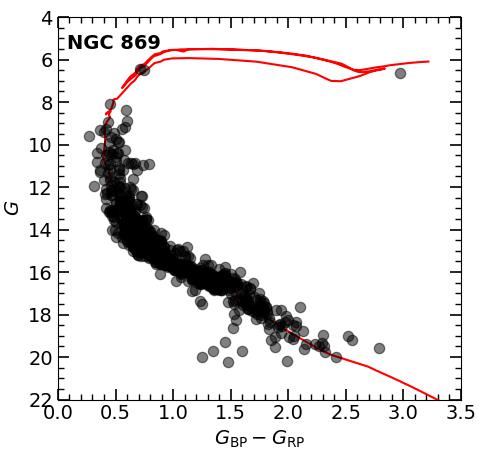}\vspace{1mm}
    \includegraphics[width=0.26\textwidth]{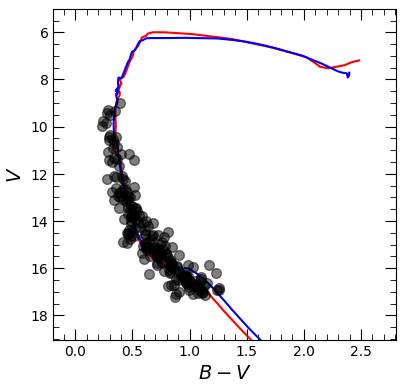}\vspace{1mm}
	\includegraphics[width=0.26\textwidth]{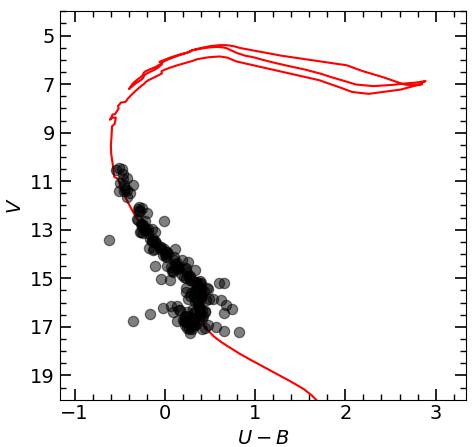}\vspace{1mm}\\    
	\includegraphics[width=0.26\textwidth]{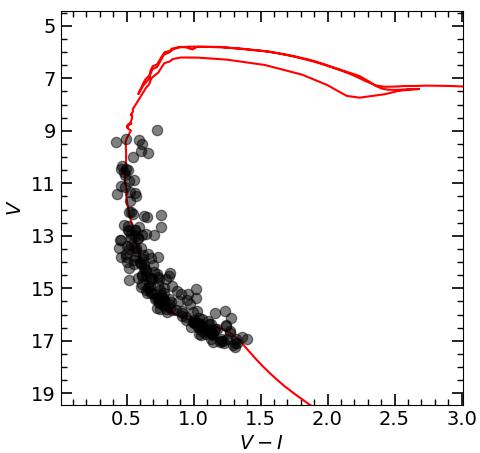}\vspace{1mm}
	\includegraphics[width=0.26\textwidth]{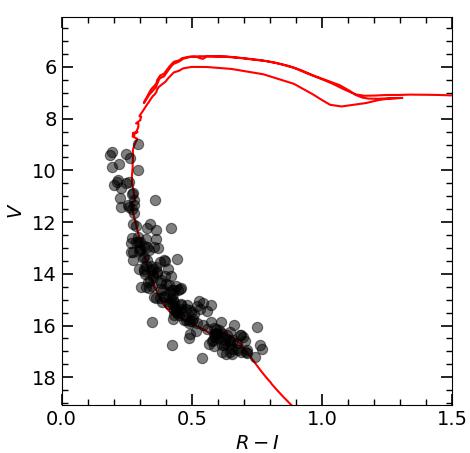}\vspace{1mm}
	\caption{The $G-(G_{BP}-G_{RP})$ (830 members), $V-(B-V)$ (193 members), $V-(U-B)$ (180 members), $V-(V-I)$ (192 members) and $V-(R-I)$ (191 members) CMDs of NGC~869. The solid red curves show the best-fitting non-rotating PARSEC isochrone corresponding to the adopted reddening, distance modulus, metallicity, and age listed in Table~\ref{lit-comp}, while the blue solid curve in the $V-(B-V)$ CMD represents an isochrone with the same parameters but solar metallicity ($Z=0.0152$).}
	\label{fig:ngc869-cmdubvri}
\end{figure*}

\begin{figure*}
	\centering{
    	\includegraphics[width=0.26\textwidth]{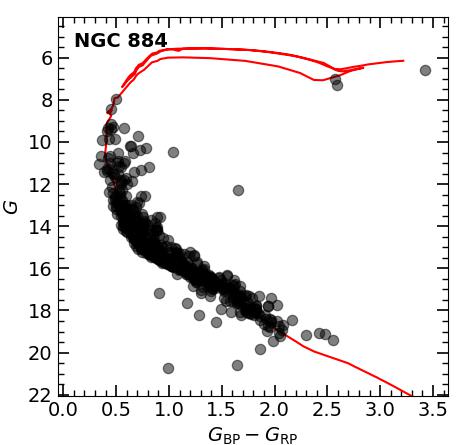}\vspace{1mm}
		\includegraphics[width=0.26\textwidth]{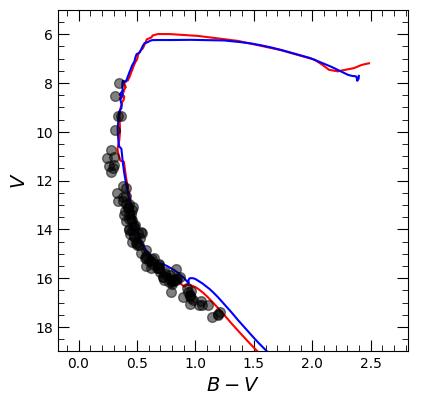}\vspace{1mm}
		\includegraphics[width=0.26\textwidth, height=0.245\textwidth]{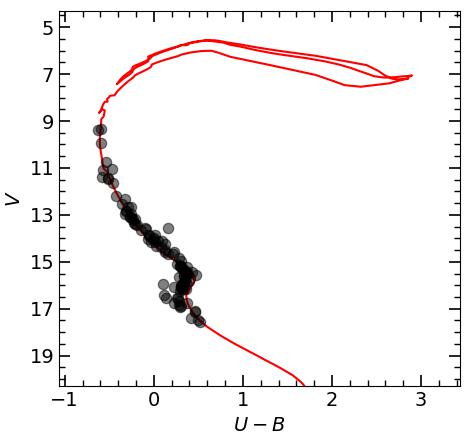}\vspace{1mm}\\
		\includegraphics[width=0.26\textwidth]{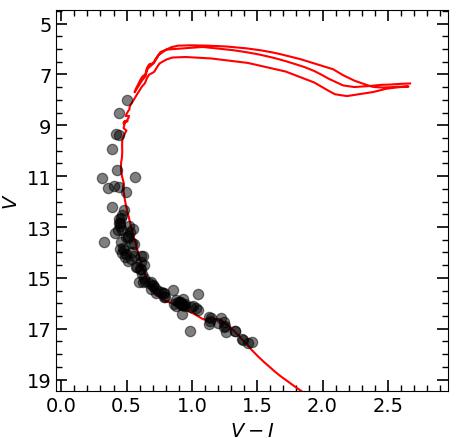}\vspace{1mm}
		\includegraphics[width=0.265\textwidth]{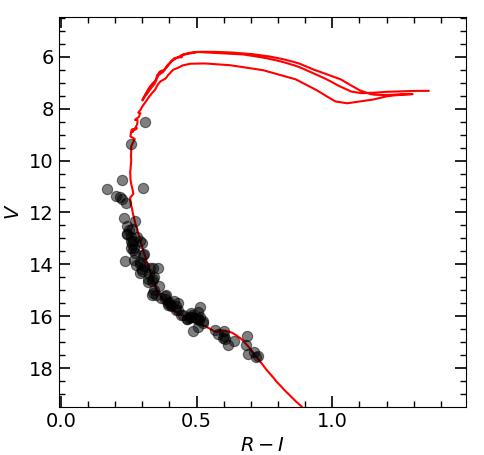}\vspace{1mm}
		\caption{The $G-(G_{BP}-G_{RP})$ (644 members), $V-(B-V)$ (104 members), $V-(U-B)$ (100 members), $V-(V-I)$ (104 members) and $V-(R-I)$ (99 members) CMDs of NGC 884. The solid red curves show the best-fitting non-rotating PARSEC isochrone corresponding to the adopted reddening, distance modulus, metallicity, and age listed in Table~\ref{lit-comp}, while the blue solid curve in the $V-(B-V)$ CMD represents an isochrone with the same parameters but solar metallicity ($Z=0.0152$).}
		\label{fig:ngc884-cmdubvri}
	}
\end{figure*}

The final isochrone fits to the $V$ versus $(B-V)$, $(U-B)$, $(R-I)$, $(V-I)$, and $G$ versus $(G_{\rm BP}-G_{\rm RP})$ CMDs for both clusters are presented in Figures~\ref{fig:ngc869-cmdubvri} and \ref{fig:ngc884-cmdubvri}. The quoted age and distance uncertainties correspond to the 68\% credible intervals obtained from the marginalised Bayesian posterior and therefore describe the statistical uncertainty of the fit under the adopted metallicity, reddening, and stellar-evolution models. Systematic effects such as unresolved binaries, stellar rotation, and model-dependent uncertainties are not explicitly incorporated into these formal errors and may contribute additional uncertainty to the absolute age scale. The sensitivity tests presented above, however, show that the effects of differential reddening, photometric offsets, and the adopted metallicity on the derived parameters are limited within the range explored.

The parallax-based distances listed in Table~\ref{tab-PM} are somewhat larger than the CCD-based distances of approximately 2300 pc. Such a difference is not unexpected, since converting parallaxes into distances is non-trivial and can lead to asymmetric uncertainties and non-Gaussian distance distributions \citep{hunt2024}. In addition, \textit{Gaia} parallaxes are subject to systematic effects that depend on source properties and position on the sky \citep{lindegren21}. We therefore retain the CCD isochrone-fitting distances as our primary values.

To assess the sensitivity of the CMD fit to the adopted metallicity, we repeated the Bayesian fitting at solar metallicity, $Z=0.0152$, while keeping the adopted reddening values fixed. The solar-metallicity fits yield MAP values of $\log(t/{\rm yr})=7.20$ ($\sim15.8$ Myr) and $7.12$ ($\sim13.1$ Myr), with distance moduli of $11.74$ and $12.00$ mag for NGC~869 and NGC~884, respectively. These can be compared with the adopted $Z=0.008$ solutions of $\log(t/{\rm yr})=7.08$ ($\sim12$ Myr), with distance moduli of $\sim 11.80$ and $\sim 11.82$ mag. Thus, the solar-metallicity assumption changes the derived age by $\sim3.8$ Myr for NGC~869 and $\sim1.1$ Myr for NGC~884, while the distance modulus changes by only $0.06$ and $0.18$ mag, respectively. Although the differences between the two metallicity assumptions are relatively small in the CMDs, the $Z=0.008$ isochrone (red) provides a better overall match to the observed main sequence than the solar-metallicity isochrone (blue), particularly for NGC~884 (Figure~\ref{fig:ngc869-cmdubvri} and~\ref{fig:ngc884-cmdubvri}). These tests show that the inferred parameters have some dependence on the adopted metallicity, but that the $Z=0.008$ solution provides the preferred match to the observations.

\begin{figure}
	\begin{center}
		\includegraphics[width=0.9\linewidth]{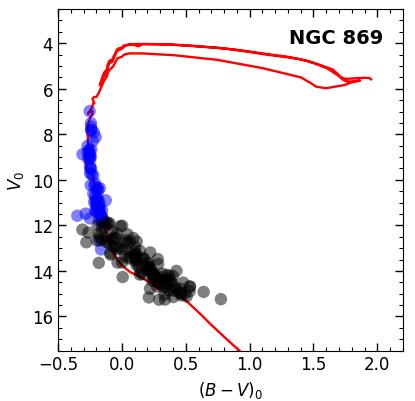} 
		\includegraphics[width=0.9\linewidth]{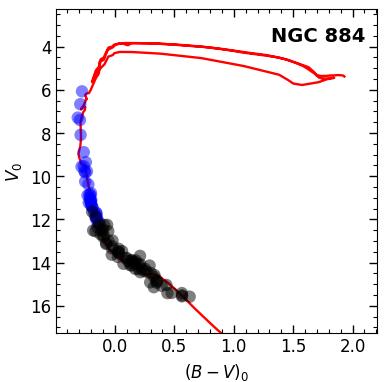}
		\caption{Dereddened $V_0$ versus $(B-V)_0$ CMDs of NGC~869 and NGC~884 after applying the star-by-star differential-reddening correction. Blue symbols indicate the early-type probable members used as reddening tracers, while grey symbols represent the remaining probable members. The red curves show the same B12 isochrones used in the original fitting, with the adopted age, metallicity, and distance modulus. The corrected CMDs show a somewhat narrower main sequence, particularly for the early-type members, while the adopted isochrone parameters remain unchanged.}
		\label{fig:dered}
	\end{center}
\end{figure}

We then examined the effect of spatial reddening variations on the derived astrophysical parameters by applying a star-by-star differential-reddening correction to the probable members. The individual $E(B-V)$ values derived for the early-type members using the Q-method were used as reddening tracers, and local reddening values for the remaining members were estimated by spatial interpolation at their positions, following approaches adopted in previous studies \citep{sharma07,hur12,lim13,lim14,sung13b}. The corresponding extinction and colour-excess corrections were then applied to obtain the dereddened $V_0$ and $(B-V)_0$ values. The resulting dereddened CMDs are shown in Figure~\ref{fig:dered}. The differential-reddening correction slightly reduces the width of the main sequence, particularly among the early-type members, indicating that spatial reddening variations contribute to the observed CMD scatter. Nevertheless, the same isochrone parameters provide a consistent fit to the corrected CMDs, with an age of 12~Myr, $[\mathrm{Fe/H}]=-0.30$, and $(V-M_V)_0=11.80$. This demonstrates that the derived astrophysical parameters are robust against the observed spatial reddening variations and that the adopted global reddening derived from the early-type members is adequate for their determination.

\begin{table*}
	\centering
	\caption{Astrophysical parameters of NGC~869 and NGC~884 compared with values reported in previous studies. The listed parameters are colour excess, true distance modulus, photometric distance, metallicity, age, and number of members.}
	\label{lit-comp}
	\begin{tabular}{lcccccccc}
		\hline
		\multicolumn{8}{c}{\textbf{NGC 869}} \\
		\hline
		$E(B-V)$ & $(V-M_V)_0$ & $d$ (kpc) & [Fe/H] & Age (Myr) & $N$ & Photometry & Reference \\
		\hline
		$0.62\pm0.10$& $11.80\pm0.22$ & $2.29\pm0.23$ & $-0.30 \pm 0.10$ & $12\pm1$ & 198 & CCD UBVRI & This paper \\ 
		\hline
		$0.51$ & -- & $2.28$ & -- & $18.81$ & $403$ & \textit{Gaia} DR3 & \citep{hunt23} \\
		$0.51 \pm 0.02$ & -- & -- & -- & -- & $10$ & Spectroscopy & \citep{yu23}  \\
		$0.55$ & -- & $2.50$ & Half solar & $15.14$ & $503$ & \textit{Gaia} DR2 & \citep{Song2022} \\
		$0.53$ & -- & $2.35^{+0.18}_{-0.20}$ & $-0.29$ & $20.00$ & -- & \textit{Gaia} EDR3 & \citep{Dalessandro2021}  \\
		$0.56\pm0.01$ & -- & $2.25 \pm 0.06$ & $-0.10 \pm 0.08$ & $12.9 \pm 1.2$ & $705$ & \textit{Gaia} DR2 & \citep{dia21}  \\
		-- & $11.71$ & $2.20$ & -- & $15.14$ & $497$ & \textit{Gaia} EDR3 & \citep{Poggio2021} \\
		$0.55$ & $11.71$ & $2.20$ & -- & $15.10$ & $503$ & \textit{Gaia} DR2 & \citep{cantat2020} \\
		$0.58$ & -- & $2.34$ & $-0.25 \pm 0.18$ & $11.75$ & -- & \textit{Gaia} DR2 + LAMOST & \citep{zhong20}\\
		$0.48^{+0.05}_{-0.05}$ & $11.69$ & $2.18$ & solar & $14.00$ & $828$ & \textit{Gaia} DR2 & \citep{Li2019} \\
		$0.65$ & -- & $2.34$ & solar & $14.00$ & -- & \textit{Gaia} DR2 & \citep{zhong19} \\
		$0.52$ & $11.81$ & & -- & $19.05$ &  & WISE & \citep{granada18} \\
		-- &-- & $2.30$ & -- & $19.05$ & $498$ & PTF/iPTF + 2MASS & \citep{yu18} \\
		0.52 &-- & $2.40\pm0.66$ & -- & $17.4\pm1.6$ &  & IPHAS & \citep{Dambis2017} \\
		-- &-- & $2.35$ & -- & $10.80$ & $238$ & WEBDA & \citep{zhai2017} \\
		0.52 &-- & $2.30$ & -- & $19.05$ & & PPMXL + 2MASS & \citep{kharchenko16} \\
		-- &-- & $2.08$ & $-0.30$ & $12.00$ & & Spectroscopy & \citep{marsakov16} \\
		$0.55$ & $11.80 \pm 0.08$ & $2.29^{+0.09}_{-0.08}$ & -- & $13.5 \pm 1.5$ & $4808$ & CCD + 2MASS + Spitzer & \citep{currie10} \\
		$0.52 \pm 0.07$ & -- & $2.30$ & -- & $15.00$ & -- & CCD UBVRI + Spectroscopy & \citep{brag05} \\
		$0.56 \pm 0.01$ & $11.85 \pm 0.05$ & $2.30^{+0.20}_{-0.40}$ & -- & $12.8 \pm 1.0$ & -- & CCD UBVRI + Spectroscopy & \citep{slesnick2002} \\
		$0.63 \pm 0.03$ & $11.70 \pm 0.10$ & $2.20 \pm 0.10$ & -- & $12.59 \pm 1.0$ & $207$ & CCD UBVRI & \citep{capilla02} \\
		$0.54 \pm 0.02$ & $11.75 \pm 0.05$ & $2.24 \pm 0.03$ & -- & $12.59 \pm 1.26$ & $--$ & CCD UBVI & \citep{keller01} \\
		\hline
		\multicolumn{8}{c}{\textbf{NGC 884}} \\
		\hline
		E(B-V) & (V-M$_v$)$_0$ & d(kpc) & [Fe/H] & Age (Myr) & N & Photometry & reference \\
		\hline
		$0.60\pm0.08$& $11.82\pm0.21$ & $2.29\pm0.22$ & $-0.30 \pm 0.10$ & $12\pm1$ & 104 & CCD UBVRI & This paper \\
		\hline
		$0.51$ & -- & $2.28$ & -- & $18.81$ & $403$ & \textit{Gaia} DR3 & \citep{hunt23} \\
		$0.51 \pm 0.01$ & -- & -- & -- & -- & $8$ & Spectroscopy & \citep{yu23} \\
		$0.56$ & -- & $2.50$ & Half solar & $17.78$ & $322$ & \textit{Gaia} DR2 & \citep{Song2022} \\
		$0.53$ & -- & $2.35^{+0.18}_{-0.20}$ & $-0.29$ & $20.00$ & -- & \textit{Gaia} EDR3 & \citep{Dalessandro2021}  \\
		$0.55\pm0.01$ & -- & $2.15 \pm 0.01$ & $-0.08 \pm 0.09$ & $15.3 \pm 1.8$ & $475$ & \textit{Gaia} DR2 & \citep{dia21}  \\
		-- & $11.64$ & $2.13$ & -- & $17.78$ & $318$ & \textit{Gaia} EDR3 & \citep{Poggio2021} \\
		$0.56$ & $11.64$ & $2.13$ & -- & $17.80$ & $503$ & \textit{Gaia} DR2 & \citep{cantat2020} \\
		$0.56$ & -- & $2.34$ & $-0.19 \pm 0.07$ & $15.85$ & -- & \textit{Gaia} DR2 + LAMOST & \citep{zhong20}\\
		$0.52^{+0.06}_{-0.04}$ & $11.81$ & $2.31$ & solar & $14.00$ & $720$ & \textit{Gaia} DR2 & \citep{Li2019} \\
		$0.65$ & -- & $2.34$ & solar & $14.00$ & -- & \textit{Gaia} DR2 & \citep{zhong19} \\
		$0.56$ & $11.85$ & & -- & $15.85$ &  & WISE & \citep{granada18} \\
		-- &-- & $2.35$ & -- & $15.85$ & $253$ & PTF/iPTF + 2MASS & \citep{yu18} \\
		0.50 &-- & $2.30\pm0.28$ & -- & $21.88\pm1.32$ &  & IPHAS & \citep{Dambis2017} \\
		-- &-- & $2.08$ & -- & $11.7$ & $247$ & WEBDA & \citep{zhai2017} \\
		0.56 &-- & $2.35$ & -- & $15.85$ & & PPMXL + 2MASS & \citep{kharchenko16} \\
		-- &-- & $2.94$ & $-0.30$ & $13.00$ & & Spectroscopy & \citep{marsakov16} \\
		$0.52$ & $11.85 \pm 0.08$ & $2.34^{+0.09}_{-0.09}$ & -- & $14.00 \pm 1.00$ & $4993$ & CCD + 2MASS + Spitzer & \citep{currie10} \\
		$0.52 \pm 0.07$ & -- & $2.30$ & -- & $20.00$ & -- & CCD UBVRI + Spectroscopy & \citep{brag05} \\
		$0.56 \pm 0.01$ & $11.85 \pm 0.05$ & $2.30^{+0.20}_{-0.40}$ & -- & $12.8 \pm 1.0$ & -- & CCD UBVRI + Spectroscopy & \citep{slesnick2002} \\
		$0.54 \pm 0.04$ & $11.70 \pm 0.10$ & $2.20 \pm 0.10$ & -- & $12.59 \pm 1.00$ & $213$ & CCD UBVRI & \citep{capilla02} \\
		$0.54 \pm 0.02$ & $11.75 \pm 0.05$ & $2.24 \pm 0.03$ & -- & $12.59 \pm 1.26$ & $--$ & CCD UBVI  & \citep{keller01} \\
		\hline
	\end{tabular}
	\begin{tabular}{p{18cm}}
        \cite{slesnick2002} reported a total of 4528 members for the binary open cluster based on CCD~UBVRI photometry and 196 members from spectroscopic analyses. \cite{brag05}, \cite{zhong19}, and \cite{keller01} adopted samples of 1640, 2186, and 17 315 stars, respectively, for the binary cluster. \cite{yu23} derived extinction using stars with available spectroscopic data; accordingly, the number of members given in column $N$ in this study represents only those stars used in their spectroscopic analysis.
	\end{tabular}
\end{table*}

Finally, we tested the effect of the systematic photometric offsets relative to \citet{keller01}, particularly in $(U-B)$ and $(V-I)$, on the derived astrophysical parameters. Applying these offsets to the photometry changes the derived ages by at most 0.3 Myr, well within the adopted 1 Myr uncertainty, while the distance moduli change by up to 0.16 mag. Thus, the offsets relative to \citet{keller01} have a negligible effect on the age estimates and a relatively small effect on the distance estimates.

The adopted colour excesses, distance moduli ($V_0-M_V$), corresponding distances (pc), ages (Myr), and numbers of members ($N$) are listed in Table~\ref{lit-comp}. The literature values compiled in Table~\ref{lit-comp} show some dispersion in the derived astrophysical parameters of NGC~869 and NGC~884. The reported reddening values generally fall within the range $E(B-V)\simeq0.50$--$0.65$ mag, and our estimates of  $0.62\pm0.10$ for NGC 869 and $0.60\pm0.08$ mag for NGC 884 lie comfortably within the bounds of previous determinations. The reported distances also show a noticeable spread, from approximately 2.1 to 2.5 kpc, reflecting differences in the adopted photometric datasets, membership selections, and distance-estimation methods. Our photometric distances of $2.29\pm0.23$ and $2.29\pm0.22$ kpc are intermediate among these estimates and are broadly consistent with several recent determinations. The largest dispersion is seen in the reported ages, which range from approximately 11 to 20 Myr. This variation can partly be attributed to differences in the adopted reddening and metallicity, as well as to differences in membership selection and isochrone-fitting procedures. In particular, several studies assume solar or half-solar metallicity, whereas available spectroscopic studies indicate a subsolar abundance of approximately $[\mathrm{Fe/H}]\simeq-0.3$. Our adopted metallicity is based on the spectroscopic determinations, and the quantitative fitting and sensitivity tests presented in this work provide an internally consistent determination of the cluster ages and distances.

\section{Kinematics and dynamical evolution }\label{sec:dynamics}

While earlier tangential velocity analyses debated bound versus disruptive pathways (e.g., \cite{uribe02}), the interpretation of the dynamical state of the Double Cluster remains somewhat ambiguous in recent Gaia-era literature. Several studies emphasize an origin as part of a larger star-forming complex, such as the LISCA I primordial group where similarities arise from co-formation within the same molecular cloud rather than a stable bound orbit (e.g., \cite{dalessandro:2021, juan22}). A comparison of recent works reveals a moderate range in adopted cluster parameters—such as total masses (ranging from ~2,700 to over 4,500 solar masses depending on completeness limits) and mean radial velocities (e.g., \cite{tarricq:2021} alongside \cite{fu22, zhong20}). These noticeable literature discrepancies in masses, velocities, and orbital binding criteria highlight the need for a comprehensive 6D phase-space reassessment.

To investigate the present-day state of the Double Cluster, we analyse the kinematic and phase-space properties of NGC~869 and NGC~884. We combine high-precision astrometry from \textit{Gaia}~DR3 with spectroscopic radial velocities. Proper motions are used to characterise the tangential kinematics of the clusters, while radial velocities complete the 3D velocity information required to explore their phase-space structure and likely evolution. In this section, we present an analysis of the observed kinematics and the likely dynamical evolution of the binary cluster, with the aim of assessing its boundedness.

\subsection{Kinematic analysis}

\begin{figure*}
	\centering
		\includegraphics[width=\textwidth]{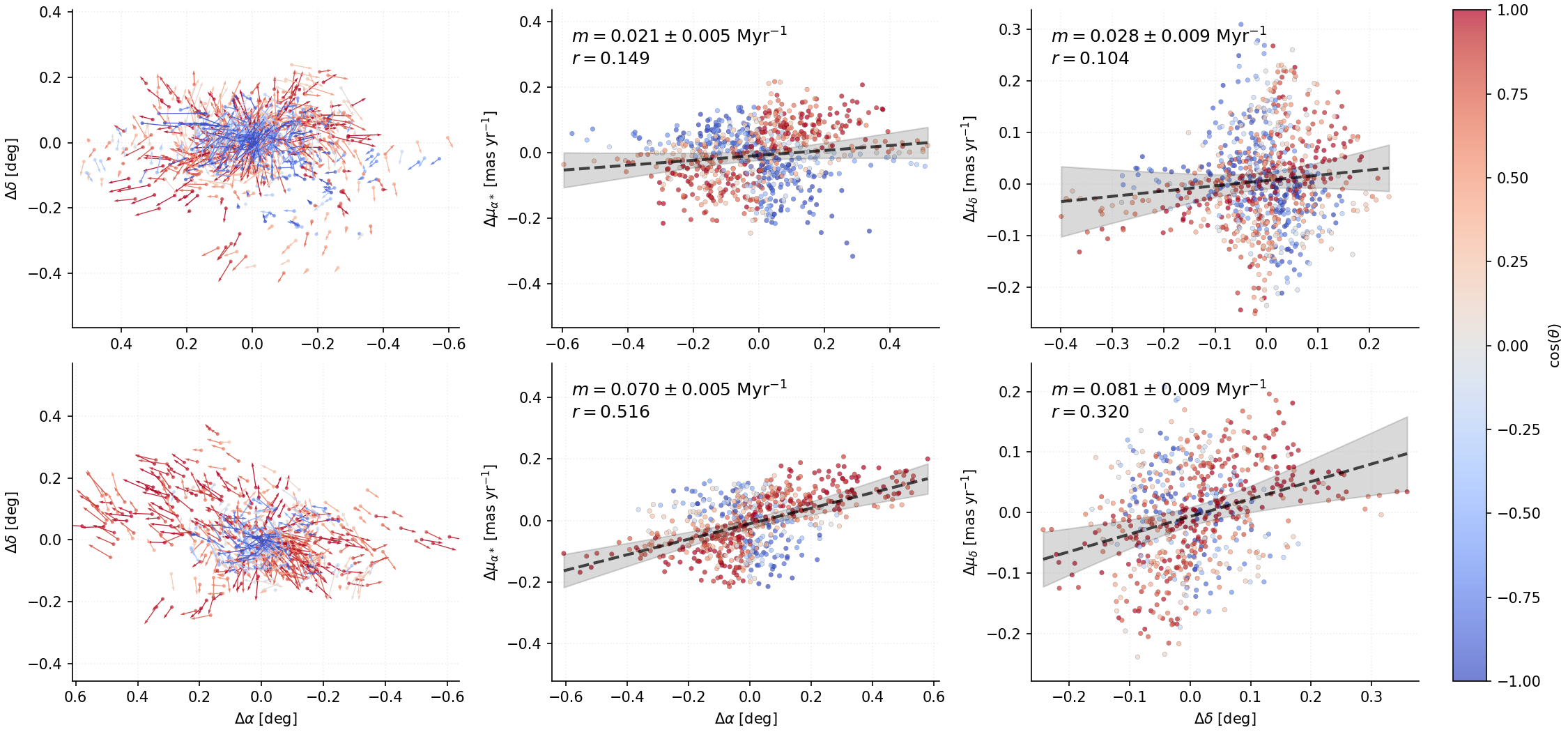}
		\caption{Internal kinematic diagnostics of the Double Cluster. The first column shows the projected stellar distribution with proper-motion vectors, while the second and third columns show $\Delta\mu_{\alpha^*}$ versus $\Delta\alpha$ and $\Delta\mu_\delta$ versus $\Delta\delta$, respectively. Dashed lines show the linear fits and grey bands their 68\% confidence intervals. The fitted slopes with $1\sigma$ uncertainties and Pearson correlation coefficients are indicated in each panel. Upper and lower rows correspond to NGC~869 and NGC~884, respectively; colours indicate $\cos(\theta)$, where $\theta$ is the angle between the proper-motion vector and the radial direction from the cluster centre.
        }
    \label{fig:binary_kin}
\end{figure*}

\begin{figure}
	\centering
		\includegraphics[width=0.5\textwidth]{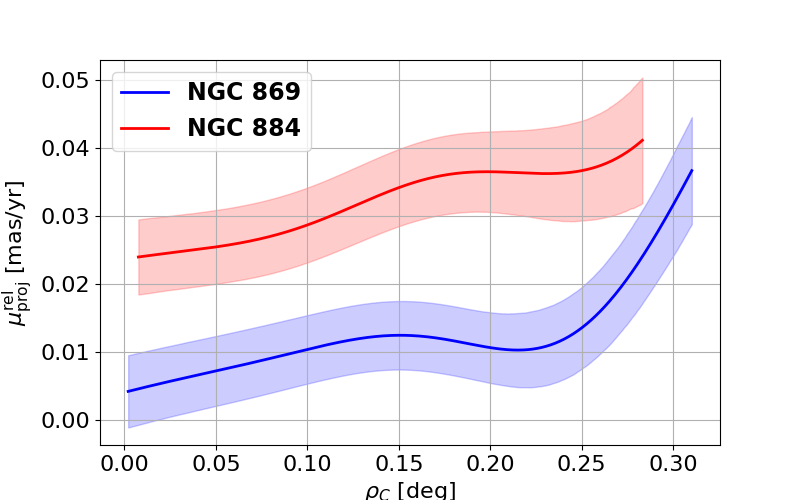}
		\caption{Profiles of the projected residual proper-motion scatter for NGC~869 (blue) and NGC~884 (red). The profiles are smoothed with a Gaussian kernel of width $\sigma = 2$~pc. The shaded bands show $\pm1$ standard error, estimated from the local variance and the number of stars contributing within the Gaussian kernel.}
    \label{fig:pm_scatter}
\end{figure}

Figure~\ref{fig:binary_kin} shows diagnostic diagrams characterising the internal velocity field of the clusters from astrometric observables. The first column displays the projected spatial distribution in $(\Delta\alpha,\Delta\delta)$ with proper-motion vectors, while the second and third columns show the proper-motion components $\Delta\mu_\alpha^*$ and $\Delta\mu_\delta$ as a function of projected position along $\Delta\alpha$ and $\Delta\delta$, respectively. Upper panels correspond to NGC~869 and lower panels to NGC~884. Linear fits highlight systematic correlations, and colours encode $\cos(\theta)$, where $\theta$ is the angle between the proper-motion vector and the radial direction from the centroid.

Both clusters show slight positive slopes in the relations between proper-motion components and projected position, although the correlations are embedded in a large scatter. For NGC~869, the fitted slopes are $m = 0.021 \pm 0.005$ Myr$^{-1}$ for $\Delta\mu_{\alpha^*}$ versus $\Delta\alpha$ and $m = 0.028 \pm 0.009$ Myr$^{-1}$ for $\Delta\mu_\delta$ versus $\Delta\delta$, with Pearson coefficients $r \simeq 0.149$ and $r \simeq 0.104$, respectively. NGC~884 exhibits substantially stronger coupling, with slopes of $m = 0.070 \pm 0.005$ Myr$^{-1}$ and $m = 0.081 \pm 0.009$ Myr$^{-1}$, with $r \simeq 0.52$ and $r \simeq 0.32$ for the $\Delta\mu_{\alpha^*}$ versus $\Delta\alpha$ and $\Delta\mu_\delta$ versus $\Delta\delta$ relations, respectively.

Using the quoted formal uncertainties of the linear fits, the fitted gradients differ from zero at approximately $4.2\sigma$ and $3.1\sigma$ for NGC~869, compared with $14.0\sigma$ and $9.0\sigma$ for NGC~884, for the $\Delta\mu_{\alpha*}$--$\Delta\alpha$ and $\Delta\mu_{\delta}$--$\Delta\delta$ relations, respectively. More directly, the differences between the NGC~884 and NGC~869 slopes are $\Delta m_{\alpha^*}=0.049\pm0.007~{\rm Myr}^{-1}$ and $\Delta m_{\delta}=0.053\pm0.013~{\rm Myr}^{-1}$, corresponding to formal significances of approximately $6.9\sigma$ and $4.2\sigma$. Together with the larger Pearson correlation coefficients reported above for NGC~884, these results quantitatively support a stronger and more coherent projected expansion signature in NGC~884, although substantial scatter remains.

A positive gradient in these projections indicates that stars at larger projected distances from the cluster centre exhibit larger outward tangential velocities, a kinematic signature commonly interpreted as cluster expansion \citep[e.g.][]{wright_mamajek:2018, kuhn:2019}. Although both clusters show such trends, the signal is weak and highly scattered in NGC~869, suggesting that any expansion is masked by intrinsic velocity dispersion, particularly in the central region, where the scatter of the data is greater. In contrast, NGC~884 displays relatively steeper gradients and tighter position–velocity coupling, consistent with a stronger and more coherent expansion pattern, although the expansion is not purely homologous.


The combination of enhanced central velocity dispersion and outward expansion at larger radii is consistent with a scenario in which a relatively bound core is embedded within an expanding halo, likely as a consequence of early residual gas expulsion \citep[e.g.][]{goodwin:2006,kruijssen:2012}. Similar enhancements in central velocity dispersion, together with evidence of simultaneous expansion and random motions, have been reported in \textit{Gaia}-based studies of young clusters and associations \citep[e.g.][]{wright_mamajek:2018,kuhn:2019}. These results support the interpretation that both components of the Double Cluster are dynamically evolving, with NGC~884 undergoing a relatively more pronounced expansion phase than NGC~869.

Figure~\ref{fig:pm_scatter} provides an independent diagnostic of the internal kinematics through the projected residual proper-motion scatter, $\mu^{\mathrm{rel}}_{\mathrm{proj}}$, as a function of projected distance from the cluster centre, $\rho_c$. In both NGC~869 and NGC~884, $\mu^{\mathrm{rel}}_{\mathrm{proj}}$ increases with radius, indicating larger relative tangential motions in the outskirts and supporting an expansion component. NGC~884 displays systematically higher scatter and a steeper, smoother rise than NGC~869, where a more stable core is likely present. Therefore, the kinematic behaviour of NGC~884 is consistent with a stronger and more coherent expansion signature, in agreement with the tighter position--velocity trends in Figure~\ref{fig:binary_kin}. The mild flattening at intermediate radii likely marks the transition from a dynamically colder inner region to a more kinematically active outer envelope. The transparent bands surrounding each smoothed profile represent the estimated statistical uncertainty at each point. They indicate the uncertainty in the median projected value at a given distance from the cluster centre, based on the Gaussian-weighted local variance and the number of contributing stars. Wider bands indicate greater uncertainty due to increased scatter or fewer contributing stars.

\subsection{Phase-space parameters}
\label{sec:phasespace}

\begin{figure*}
	\centering
	\includegraphics[width=16.5cm]{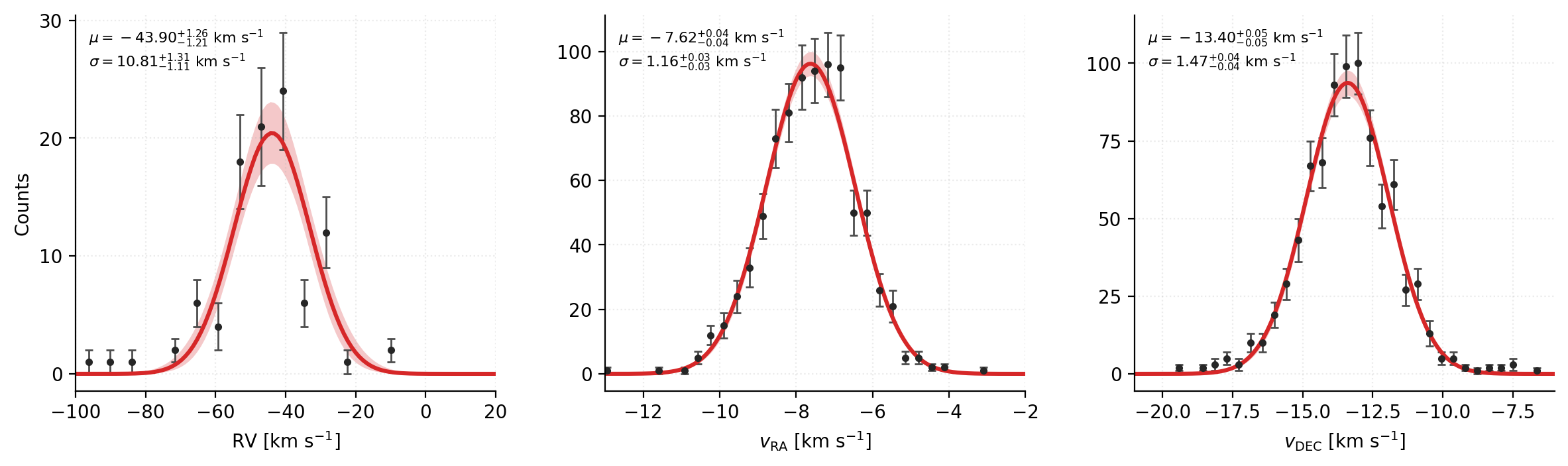}
	\includegraphics[width=16.5cm]{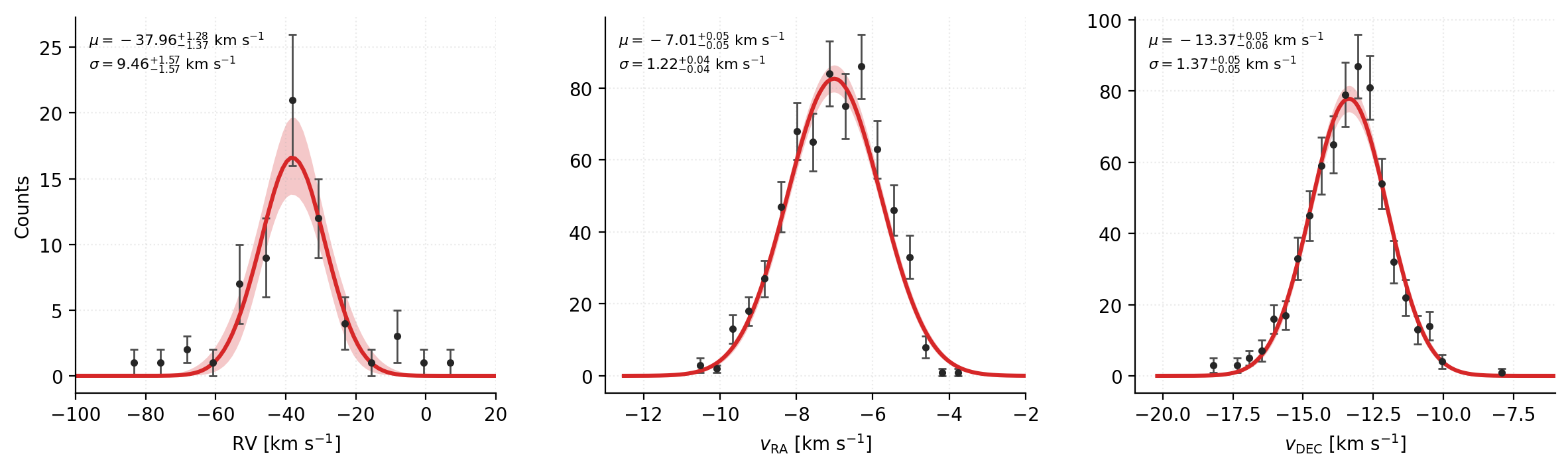}
	\caption{
    Velocity distributions for NGC~869 (upper row) and NGC~884 (lower row). The left panels show radial velocities from the LAMOST survey, while the middle and right panels show the tangential velocity components $V_{\rm RA}$ and $V_{\rm DEC}$ derived from \textit{Gaia}~DR3 proper motions and parallaxes. The red curves show the best-fitting one-dimensional Gaussian models, and the shaded regions represent the 68\% posterior credible intervals. Error bars on the binned counts represent Poisson uncertainties. The fitted mean velocities and dispersions, with their uncertainties, are indicated in each panel.
	}
	\label{fig:vel_dist}
\end{figure*}

Radial velocities for members of NGC~869 and NGC~884, obtained from the LAMOST survey \citep{zhao:2012}, were combined with proper motions and distances to complement the kinematic analysis of the system. Each velocity component (${\rm RV}$, $V_{\rm RA}$, and $V_{\rm DEC}$)\footnote{Tangential velocities were calculated using $V_{\rm RA} = 4.74\,\frac{\mu_{\alpha^*}}{\varpi}$ and $V_{\rm DEC} = 4.74\,\frac{\mu_{\delta}}{\varpi}$, with units in km s$^{-1}$.} was modelled independently with a one-dimensional Gaussian in the observed coordinate system,
\begin{equation}
\label{eq:vel_gauss}
f(v)=A\,\exp\!\left[-\frac{1}{2}\left(\frac{v-\mu}{\sigma_v}\right)^2\right],
\end{equation}
where $A$ is the amplitude, $\mu$ the mean velocity, and $\sigma_v$ the velocity dispersion. The fits were performed using the \texttt{PyMC} tool within a Bayesian framework, following the general velocity-distribution methodology adopted
by \citet{SanchezSanjuan2026}.

All stars entering the kinematic analysis were drawn from the Gaia-selected membership samples satisfying the astrometric quality criterion $\mathrm{RUWE}<1.4$ (Section~2.1). Prior to fitting the radial-velocity distributions, we applied a one-pass $5\sigma$ clipping to remove extreme RV outliers and retained only measurements satisfying $RV_{\rm err}/|RV|<0.20$, resulting in final samples of 108 stars for NGC~869 and 71 stars for NGC~884. The histogram bin widths were determined using the Freedman--Diaconis rule, which provides a robust, data-driven compromise
between noise and resolution. We did not explicitly identify or remove spectroscopic binaries. Therefore, although the astrometric and RV-quality selections reduce the influence of problematic measurements and extreme velocity values, unresolved binaries may still broaden the RV distributions and contribute to the uncertainty of the derived systemic velocities.

We note that the ${\rm RV}$ sample is smaller than the \textit{Gaia}-based tangential-velocity sample, and therefore the inferred ${\rm RV}$ moments are more sensitive to selection effects and sampling noise. Figure~\ref{fig:vel_dist} shows the radial and tangential velocity distributions for NGC~869 and NGC~884 together with the best fits.

From the Gaussian fits, we obtain systemic radial velocities of $-43.9^{+1.3}_{-1.2}~\mathrm{km\,s^{-1}}$ for NGC~869 and $-38.0^{+1.3}_{-1.4}~\mathrm{km\,s^{-1}}$ for NGC~884.
As an additional robustness check, we estimated the systemic radial velocities using the sample medians and bootstrap resampling, including the individual RV measurement uncertainties. We obtain $RV_{\rm med}=-44.97^{+1.14}_{-1.27}~{\rm km\,s^{-1}}$ for NGC~869 and $RV_{\rm med}=-37.78^{+1.97}_{-1.41}~{\rm km\,s^{-1}}$ for NGC~884. These values are consistent within the uncertainties with the Gaussian-fit centroids, indicating that the inferred systemic velocities are not strongly dependent on the assumed Gaussian form of the RV distributions.


We compare our fitted values with recent literature measurements: \citet{Soubiran:2018} (\textit{Gaia} DR2) report $-62.8\pm16.1~\mathrm{km\,s^{-1}}$ for NGC~869 based on 1 member and $-43.0\pm5.7~\mathrm{km\,s^{-1}}$ for NGC~884 based on 2 members, while \citet{hunt2024} (\textit{Gaia} DR3) find $-67.6\pm10.4~\mathrm{km\,s^{-1}}$ for NGC~869 and $-43.9\pm1.6~\mathrm{km\,s^{-1}}$ for NGC~884, based on 2 and 6 members, respectively. The larger discrepancies in these Gaia-based determinations likely reflect the small RV sample sizes. Similarly, \citet{tademir:2026} report $-38.2\pm3.2~\mathrm{km\,s^{-1}}$ (NGC~869) and $-40.2\pm3.0~\mathrm{km\,s^{-1}}$ (NGC~884) from 8 stars in total. More recently, \citet{Zhang:2026} report $-44.5\pm8.5~\mathrm{km\,s^{-1}}$ for NGC~869 (41 stars) and $-37.6\pm10.6~\mathrm{km\,s^{-1}}$ for NGC~884 (32 stars) using LAMOST medium-resolution data. Our RVs agree well with the LAMOST-based measurements of \citet{Zhang:2026} and benefit from a larger RV sample, yielding a more robust characterisation of the cluster velocity distributions.

Both clusters exhibit dynamically cold tangential components, with narrow dispersions of $\sigma_{V_{\rm RA}}\sim 1.2$~km~s$^{-1}$ and $\sigma_{V_{\rm DEC}}\sim 1.4$~km~s$^{-1}$, indicating coherent motion in the plane of the sky and similar mean tangential velocities. In contrast, the fitted mean radial velocities differ, revealing a measurable line-of-sight relative motion that contributes directly to the 3D relative velocity budget and is therefore relevant for assessing whether the pair could remain gravitationally bound. Although the ${\rm RV}$ distributions are comparatively broad, likely reflecting the smaller spectroscopic sample size, the offset in systemic ${\rm RV}$ remains a robust indicator of differential motion along the line of sight.

To compute the phase-space quantities, we transformed the observables to heliocentric Cartesian positions and velocities using \textsc{Astropy}~\citep{astropy:2022}. We adopt a right-handed Galactic convention $(X_h,Y_h,Z_h)$ with $\mathbf{X}$ pointing towards the Galactic centre, $\mathbf{Y}$ in the direction of Galactic rotation, and $\mathbf{Z}$ towards the North Galactic Pole; the heliocentric velocity components $(U_h,V_h,W_h)$ follow the same sign convention. Parameter uncertainties were propagated using Monte Carlo error propagation in the observable space. The covariance terms between the \textit{Gaia} astrometric parameters were not included in this propagation; the quoted astrometric uncertainties were sampled independently.
We also corrected velocities to the local standard of rest (LSR) by subtracting the solar motion $(U_\odot,V_\odot,W_\odot)=(11.1,12.24,7.25)\ {\rm km\,s^{-1}}$ \citep{Schoenrich:2010}, and finally transformed to a right-handed Galactocentric frame adopting a solar Galactocentric distance of $R_0=8.2\ {\rm kpc}$ \citep{mcmillan:2017}.

Table~\ref{tab:phase} summarises the phase-space centroids of NGC~869 and NGC~884 in the heliocentric Galactic Cartesian frame, expressed as $(X_h,Y_h,Z_h)$ and $(U_h,V_h,W_h)$. These values  were used to compute the relative separation and velocity of the two clusters in the binding analysis. For the subsequent N-body integrations in the Galactic potential, the phase-space coordinates were transformed to the right-handed Galactocentric reference frame.

\begin{table}
	\centering
	\caption{ Phase-space centroids of NGC~869 and NGC~884 in a heliocentric Galactic Cartesian frame. Median values are shown with associated uncertainties.}
	\label{tab:phase}
	\begin{tabular}{lcc}
		\hline
		\textbf{Parameter} & \textbf{NGC 869} & \textbf{NGC 884} \\
		\hline
		$X_h$ [kpc] & $-1.73 \pm 0.01$ & $-1.76 \pm 0.01$ \\
		$Y_h$ [kpc] & $1.75 \pm 0.01$  & $1.76 \pm 0.01$ \\
		$Z_h$ [kpc] & $-0.160 \pm 0.001$   & $-0.154 \pm 0.001$ \\
		$U_h$ [$\mathrm{km\,s^{-1}}$] & $33.8 \pm 2.1$    & $28.8 \pm 1.0$ \\
		$V_h$ [$\mathrm{km\,s^{-1}}$] & $-30.7 \pm 2.4$   & $-25.4 \pm 2.12$ \\
		$W_h$ [$\mathrm{km\,s^{-1}}$] & $-12.5 \pm 0.2$   & $-13.1 \pm 0.3$ \\
		\hline
	\end{tabular}
\end{table}

While the Double Cluster reveals non-identical motions and expansion signatures in both clusters, we use their mean spatial and kinematic properties to assess possible dynamical interaction. Based on data from Table~\ref{tab:phase}, we compute the relative velocity of NGC~884 with respect to NGC~869 and project it onto the line joining their centres. The mean resulting projected velocity is 3.8$\pm$4.0~$\mathrm{km\,s^{-1}}$, based on 10,000 Monte Carlo realisations. The positive mean value suggests that NGC~884 is receding from NGC~869 along their mutual axis with a probability of approximately 85\% (see Figure~\ref{fig:escape}, left panel).

\subsection{Mass Estimation}
\label{sec:mass_est}

\begin{figure*}
	\centering
    \includegraphics[width=0.96\textwidth]{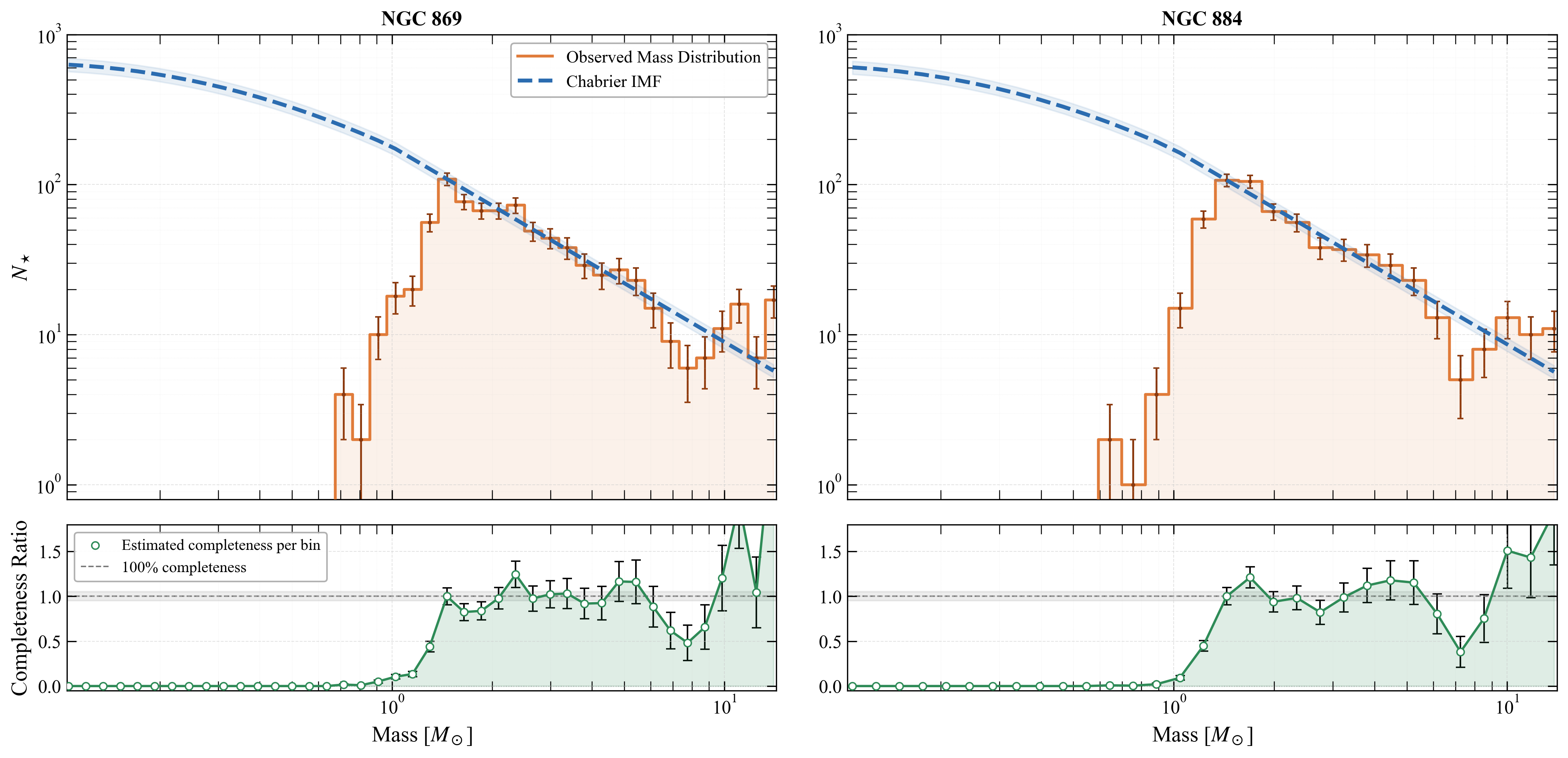}
		\caption{
    Mass distributions and completeness ratios for NGC~869 (left) and NGC~884 (right). \textit{Top panels:} observed stellar mass distributions (orange histograms) with $1\sigma$ Poisson uncertainties compared with the best-fitting \citet{chabrier:2003} mass function (blue dashed curves);
    the blue shaded regions show the formal $1\sigma$ uncertainty of the mass-function fit. \textit{Bottom panels:} completeness ratios between the observed counts and the mass-function prediction, with uncertainties estimated through Monte Carlo error propagation. The grey dashed line marks 100\% completeness level and the grey shaded region indicates a $\pm5$\% tolerance around this level.
}

    \label{fig:completeness_analysis}
\end{figure*}

The dynamical interpretation of the clusters requires reliable mass estimates. Accordingly, the stellar masses of the 830 members of NGC~869 and the 644 members of NGC~884 identified in Section~\ref{sec:membership-selec} were derived using the B12 PARSEC isochrones~\citep{bressan2012}, based on their positions in the CMD and the adopted cluster ages. We estimate observed masses of $2500 \pm 109\,M_{\odot}$ for NGC~869 and $1900 \pm 88\,M_{\odot}$ for NGC~884. These values are consistent with those reported by~\citet{tademir:2026}, namely $2119\,M_{\odot}$ and $1800\,M_{\odot}$. However, our estimates are likely biased low, particularly at the low-mass end, where faint stars are incompletely sampled as a consequence of the selection criteria adopted in this study.

Recently, \citet{Celli:2024} derived total masses of $3678\,M_{\odot}$ for NGC~869 and $2293\,M_{\odot}$ for NGC~884 after applying a broken power-law correction. These values correspond to increases of $\sim 47\%$ and $\sim 21\%$, respectively, relative to our observed mass estimates. This higher total mass has important implications for assessing the boundedness of the system.

Considering that the mass estimates reported by~\citet{Celli:2024} may still be affected by incompleteness, we applied an additional correction procedure. We fitted a~\citet{chabrier:2003} mass function (MF) to the highest-mass bins of the observed distribution, where Poisson noise is comparatively low. Using this fit, we integrated the MF from $0.1\,M_{\odot}$ to the highest observed mass to estimate the total cluster mass.

Figure~\ref{fig:completeness_analysis} shows the comparison between the fitted Chabrier MF and the observed mass distribution. In the top panels, the data appear to be approximately complete in the mass range $\sim 1.5$--$6.0\,M_{\odot}$, where the completeness level (bottom panels) is close to unity. At higher masses ($>10\,M_{\odot}$), the distribution becomes more scattered, which we attribute to small-number statistics and the increased uncertainty in mass estimates for post-main-sequence stars, where degeneracies are more pronounced. Below $\sim 1.5\,M_{\odot}$, the completeness ratio drops to zero. After integrating the fitted MF, we estimate total cluster masses of $6409^{+224}_{-200}\,M_{\odot}$ for NGC~869 and $4547^{+110}_{-91}\,M_{\odot}$ for NGC~884. These values correspond to increases by factors of 2.6 and 2.4 for
NGC~869 and NGC~884, respectively, relative to the observed mass estimates. The quoted uncertainties represent the formal uncertainties from the mass-function fitting and do not include systematic effects associated with membership selection, incompleteness, or the extrapolation of the mass function towards low stellar masses.

\begin{figure*}
	\centering
    \includegraphics[width=0.95\textwidth]{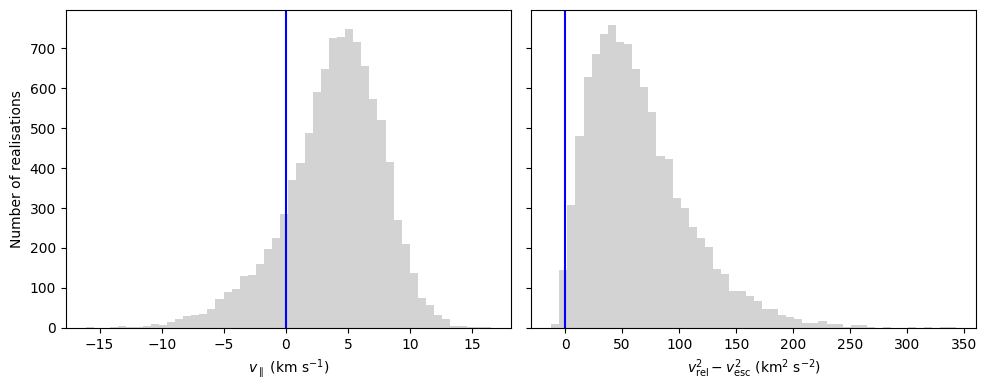}
		\caption{Projected relative velocity (left panel) and binding condition (right panel) for the Double Cluster based on the mean spatial and kinematic parameters and their uncertainties (see Table~\ref{tab:phase}) and the total mass of the system and its uncertainty (see Section~\ref{sec:mass_est}). Left panel: Distribution of the relative velocity of NGC~884 with respect to NGC~869 projected onto the line joining their centres, based on 10,000 Monte Carlo realisations. The blue vertical line marks zero projected velocity. Right panel: Distribution of $v_{\mathrm{rel}}^2-v_{\mathrm{esc}}^2$, where $v_{\mathrm{esc}}$ is the two-body escape velocity at the instantaneous separation. Negative values correspond to bound configurations. The blue vertical line indicates the zero-energy boundary.}
    \label{fig:escape}
\end{figure*}

Considering the previous result, we evaluate whether NGC~869 and NGC~884 are gravitationally bound in the simplified approximation of two point masses. Using their mean heliocentric positions and space velocities, along with their uncertainties, as evaluated in Table~\ref{tab:phase}, together with the estimated cluster masses and their uncertainties, we perform 10,000 Monte Carlo realisations, propagating the uncertainties in positions, velocities, and masses. For each realisation, we compare the squared relative velocity between cluster centroids, $v_{\rm rel}^2$, to the squared escape velocity
\begin{equation*}
    v_{\mathrm{esc}}^2 = \frac{2G(M_{tot})}{r},
\end{equation*}
where $M_{\rm tot}$ is the sum of the cluster masses, $r$ is the separation between their centres, and $G$ is the gravitational constant. 
A realisation is classified as bound when $v_{\rm rel}<v_{\rm esc}$,
equivalently when $v_{\rm rel}^{2}-v_{\rm esc}^{2}<0$. We find a very low binding probability of $P_{\mathrm{bound}} \approx 0.8\%$ (see distribution in the right panel of Figure~\ref{fig:escape}). The choice of cluster distances has only a minor effect on the dynamical analysis. The \textit{Gaia} parallax-based distances are $\sim150$--$200$ pc larger than the CCD-based values, but the similar offset for both clusters results in only a few-percent change in the mutual escape velocity. Thus, the adopted distances do not affect our conclusion that NGC~869 and NGC~884 are not gravitationally bound.

To assess the sensitivity of the binding probability to the systemic radial velocities, we defined the line-of-sight velocity difference as $\Delta RV=|RV_{884}-RV_{869}|$ and its combined uncertainty as $\sigma_{\Delta RV}=(\sigma_{RV,869}^{2}+\sigma_{RV,884}^{2})^{1/2}$, assuming independent RV uncertainties. We then repeated the Monte Carlo analysis after reducing the nominal velocity difference by $\sigma_{\Delta RV}$, thereby shifting the system towards a configuration more favourable to gravitational binding. The resulting probability of realisations satisfying $v_{\rm rel}<v_{\rm esc}$ is $P_{\rm bound}\simeq3.4\%$. Thus, a conservative $1\sigma$ reduction of the RV difference does not alter the conclusion that the Double Cluster is presently unlikely to be gravitationally bound. Although neglecting the astrometric covariance may slightly modify the propagated phase-space uncertainties, we do not expect it to affect this qualitative conclusion, given the very low nominal binding probability ($P_{\rm bound}\simeq0.8\%$) and the similarly low probability obtained in this more favourable configuration.

For a reference point, we evaluated the total mass required for a $50\%$ binding probability. Running 1,000 Monte Carlo realisations to propagate uncertainties in positions and velocities, we find the mass corresponding to a 50\% binding probability to be
\begin{equation*}
M_{50} \approx (2.6 \pm 1.2) \cdot 10^5 M_\odot,     
\end{equation*}
which exceeds the estimated combined mass of the Double Cluster by more than an order of magnitude (a factor of $\sim 24$). However, determining whether the pair is gravitationally bound requires following its evolution under the Galactic tidal field while propagating the observational uncertainties.

\subsection{Evolution of NGC~869 and NGC~884}
\label{sec:binary_evolution}

\begin{figure*}
	\centering
		\includegraphics[width=0.98\textwidth]{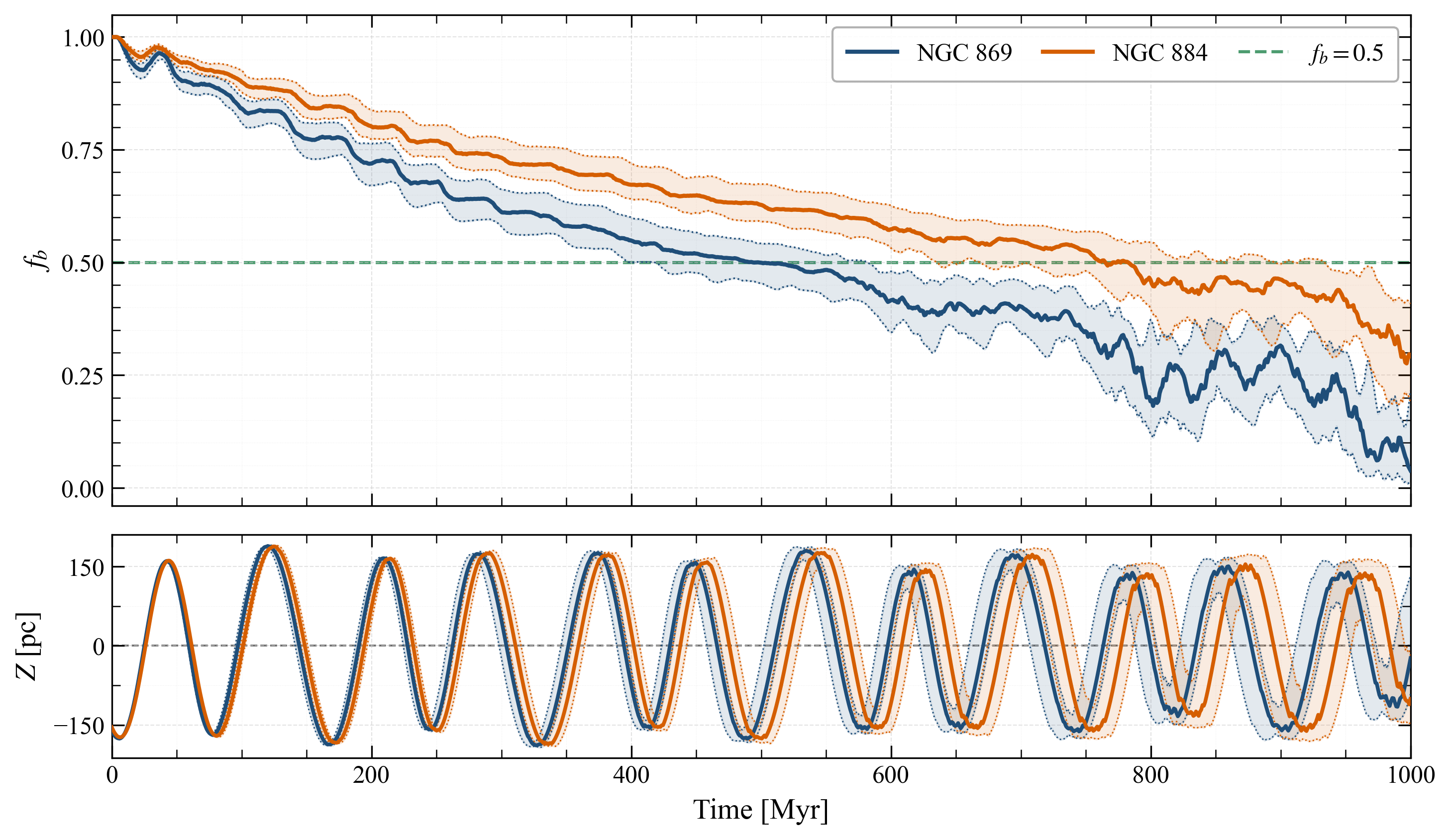}
		\caption{Time evolution of the Double Cluster components from the ensemble of 50 $N$-body realisations. \textit{Top panel:} Evolution of the bound stellar fraction for NGC~869 and NGC~884. \textit{Bottom panel:} Vertical ($z$) motion of the cluster centroids in the Galactic potential. Solid curves show the median evolution across the realisations, while the shaded regions enclose the 16th--84th percentiles. The horizontal green line marks $f_b=0.5$.
    }
    \label{fig:dissolution}
\end{figure*}

\begin{figure}
	\centering
    \includegraphics[width=0.48\textwidth]{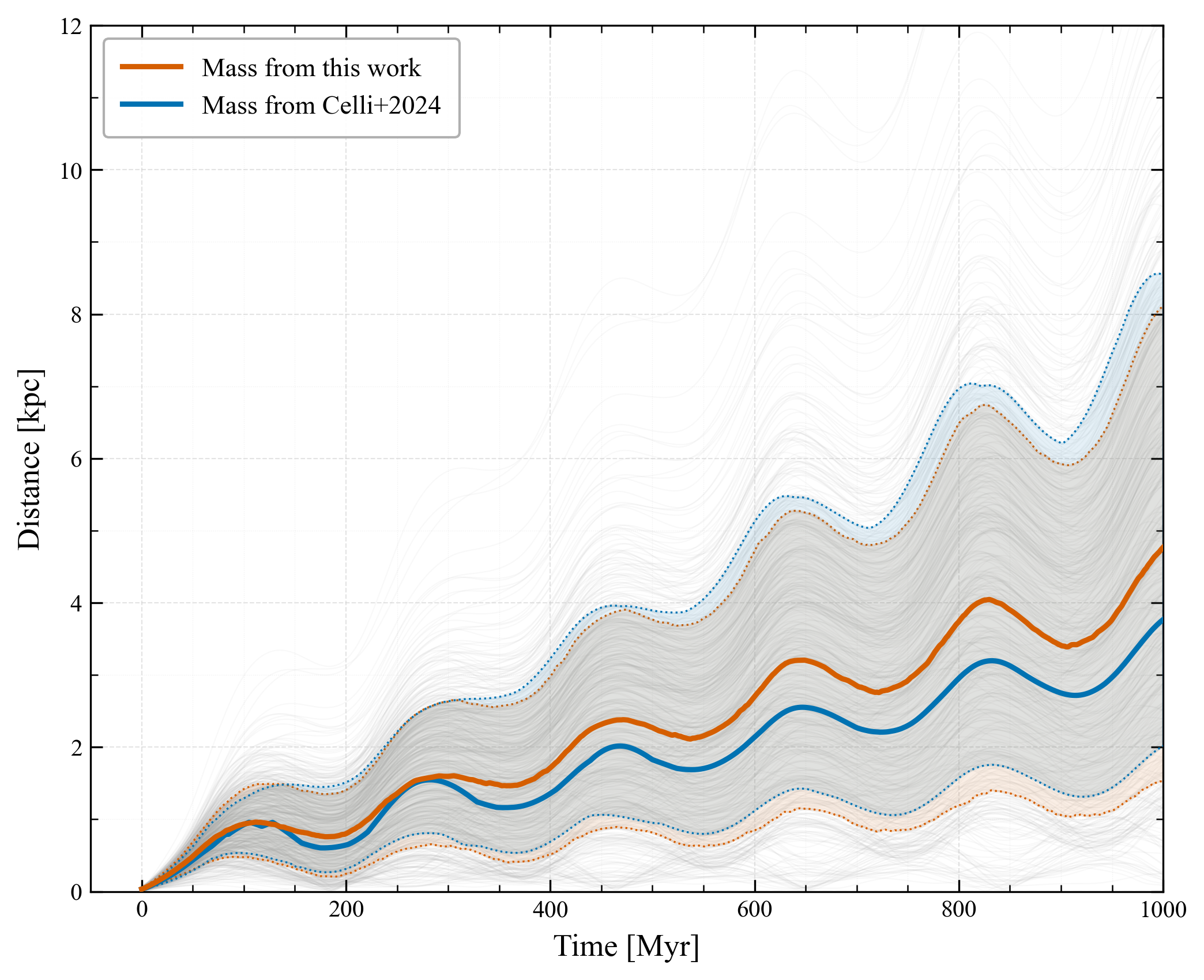}
		\caption{Evolution of the separation between the cluster centroids for the two mass prescriptions: masses from~\citet{Celli:2024} (blue) and the completeness-corrected masses derived in this work (orange). The thick curves show the median separation across the corresponding simulation ensemble, while the shaded regions enclose the 16th--84th percentile ranges. Individual realisations for the corrected-mass scenario are shown in grey.
}
    \label{fig:distance_evol}
\end{figure}

To explore the dynamical fate of the Double Cluster, we generated an ensemble of direct $N$-body simulations consistent with the present-day structural and
kinematic constraints, following the general numerical framework of
\citet{SanchezSanjuan2026} and adapting it to NGC~869 and NGC~884.

All computations were carried out using the Astrophysical MUltipurpose Software Environment (\texttt{AMUSE}), a framework for simulating astrophysical systems within a unified numerical environment \citep{pz13, pelupessy:2013, portegies_book:2018}.

The direct \(N\)-body simulations presented in this section adopt the total cluster masses reported by~\citet{Celli:2024}, which allow computationally feasible particle numbers for the full stellar realisations to assess the dissolution of the clusters. We then separately explored the impact of the larger completeness-corrected masses derived in this work using a simplified two-particle representation in order to evaluate whether the revised masses modify the inferred dynamical state and boundedness of the system.

\subsubsection{Dissolution of NGC 869 and NGC 884}

In the first set of simulations, we generated 50 realisations with initial conditions constructed as King distributions using the \texttt{AMUSE} sampler. The central concentration, defined as $c=\log_{10}(r_t/r_c)$ and derived from Table~\ref{tab:tbl-struct}, is $1.54\pm0.04$ for NGC~884 and $1.71\pm0.03$ for NGC~869. The uncertainties were estimated through Monte Carlo propagation, assuming independent Gaussian uncertainties for $r_c$ and $r_t$ and adopting the 16th--84th percentile range of the resulting $c$ distribution. Within the King formalism, these values correspond to dimensionless central potential depths of $W_0=7.04\pm0.13$ and $W_0=7.60\pm0.10$,  respectively. At the start of each simulation ($t=0$, corresponding to the present-day configuration), both clusters were assumed to be in virial equilibrium.
Stellar masses were drawn from a Chabrier IMF~\citep{chabrier:2003} up to the total estimated mass reported by~\citet{Celli:2024}, with all stars assumed to be single, as binary populations are beyond the scope of this work. Each realisation was generated through Monte Carlo propagation of the parameter uncertainties, including the centroid positions and total velocity vectors listed in Table~\ref{tab:phase}.

We evolved the clusters self-consistently within \texttt{AMUSE} by coupling the direct $N$-body integrator \texttt{ph4} with the stellar-evolution module \texttt{SeBa} for 1 Gyr. The \texttt{ph4} code employs a fourth-order Hermite scheme with block time-steps optimised for star-cluster dynamics \citep{mcmillan_ph4:2012}, while \texttt{SeBa} accounts for stellar mass loss and remnant formation through parametrised stellar evolution \citep{portegies_seba:1996}. The clusters were evolved in an axisymmetric Galactic potential following \citet{allen:1991}, scaled to a solar circular speed of $v(R_0)=232.8~\mathrm{km\,s^{-1}}$ \citep{mcmillan:2017}. Integrations used a fixed time-step of 0.1~Myr.

From our simulations, we track the likely bound stellar component in each system through the bound fraction, $f_b \equiv N/N_0$. For both clusters, we first determined the centre-of-mass position and systemic velocity. We then computed the kinetic energy in the cluster-centric frame, $E_K = \tfrac{1}{2}\,m\,\lvert \boldsymbol{v} - \boldsymbol{v}_{\rm c} \rvert^2$ and the gravitational potential energy, $E_P$, due to the stellar system. The total energy was defined as $E_T = E_K + E_P$, where stars with $E_T < 0$ are classified as gravitationally bound. This criterion allows us to follow the time evolution of the bound stellar population in a physically self-consistent way.

The top panel of Figure~\ref{fig:dissolution} shows the statistical evolution of $f_b$, indicating that both components of the Double Cluster dissolve progressively in the Galactic environment. NGC~869 dissolves faster than NGC~884, reaching $f_b = 0.5$ in $\sim 500$~Myr, whereas NGC~884 reaches the same bound fraction after $\sim 750$~Myr. The systematically steeper decline of NGC~869 suggests that, for the present-day structural and orbital parameters, it is more susceptible to disruption despite its larger initial bound mass. The continued decrease of $f_b$ towards the end of the integrations indicates that the dissolution of both clusters proceeds beyond the $1$~Gyr interval explored here. Because the clusters are initialised in virial equilibrium, the corresponding survival times should be regarded as conservative upper estimates under the adopted initial conditions.

This dissolution is driven not only by internal two-body relaxation but also by external tidal heating \citep{gnedin:1997}. As the clusters orbit within the disc, repeated passages through the Galactic plane can produce disc shocking, injecting energy into the system, increasing the velocity dispersion, and stripping stars from the outskirts \citep{Ostriker:1972,aguilar:1988, gieles:2008, webb:2014}. The bottom panel of Figure~\ref{fig:dissolution} shows the vertical ($z$) centroid motions of both clusters, which remain coherent and oscillatory, with amplitudes $|z| \sim 150$--200~pc, as expected in the Galactic disc potential \citep{binney:2008}.

The $z$-evolution further indicates that the oscillations gradually dephase over the forward integration, consistent with increasingly independent evolution of the two clusters under the adopted present-day conditions.



\subsubsection{Fate of the Double Cluster}

To explore the impact of the completeness-corrected masses derived in this work, we repeated the dynamical analysis using an alternative setup. While the direct \(N\)-body realisations based on the masses reported by~\citet{Celli:2024} required \(\sim13{,}000\) particles, adopting the revised masses would increase the particle number to more than \(30{,}000\), resulting in a substantial computational cost.
To mitigate this, we performed a suite of 500 simulations using a simplified two-particle representation, in which each particle carries the total mass of one cluster. The initial conditions were randomly sampled according to the observational uncertainties in positions, velocities, and masses. However, we note that these mass uncertainties are strictly formal and do not include systematic effects associated with membership selection, incompleteness, or mass-function extrapolation. Nevertheless, because the mass required for a $50\%$ binding probability exceeds the combined mass of the system by a factor of $\sim 24$, such systematic shifts are far too small to alter the dynamical conclusion. Using the particle positions, we computed the temporal evolution of their separation in order to assess whether the different mass-corrected scenarios lead to distinct conclusions regarding the boundedness of the binary system. We also performed the same analysis using the observed masses derived directly from the identified members, ($2500\pm109\,M_{\odot}$ for NGC~869 and $1900\pm88\,M_{\odot}$ for NGC~884). This additional test likewise yields an unbound configuration, with the clusters continuing to separate. Thus, the observed, literature, and completeness-corrected mass estimates all lead to the same qualitative conclusion that the Double Cluster is not gravitationally bound.

Figure~\ref{fig:distance_evol} shows the evolution of the centroid separation as a function of time, comparing the results obtained using the masses reported by~\citet{Celli:2024} (based on an ensemble of 50 simulations) with those derived from our corrected mass estimates. In both scenarios, the centroid separation increases rapidly during the first \(\sim100\,\mathrm{Myr}\), reaching distances of approximately \(1\,\mathrm{kpc}\), and continues to grow throughout the integration. 
The rapid increase in separation during the simulations is consistent with ballistic divergence driven by the present-day relative motion, while the subsequent growth is probably enhanced by Galactic shear
 \citep[e.g.][]{binney:2008,portegies:2010}. Superimposed on this secular trend, the separation exhibits periodic oscillations associated with differences in the Galactic orbits of the clusters.

Beyond \(\sim100\,\mathrm{Myr}\), different mass prescriptions produce a qualitatively similar evolution of the cluster separation, with largely overlapping uncertainty envelopes. The simulations show that the two clusters continue to diverge throughout the forward integrations. Combined with the very low present-day binding probability ($P_{\rm bound}\approx0.8\%$), this indicates that the Double Cluster is currently not gravitationally bound. The two clusters therefore evolve independently under the influence of the Galactic
potential \citep{portegies:2010}. This behaviour is consistent with the present relative velocity exceeding the mutual escape velocity, as found in other young cluster pairs and associations \citep[e.g.][]{baumgardt:2007,gavagnin:2016,pang:2021_2,juan22}.

\section{Conclusions}
\label{sec:conclusions}

We revisited the nature and dynamical state of the Double Cluster using new CCD $UBVRI$ photometry, \textit{Gaia} DR3 astrometry, LAMOST radial velocities, mass estimates, and dynamical modelling. Although NGC~869 and NGC~884 have closely similar astrophysical properties, their present-day kinematics and binding analysis indicate that they do not form a gravitationally bound binary cluster. Their similarity nevertheless remains consistent with a common origin within the same larger star-forming environment.

Using HDBSCAN in the five-dimensional \textit{Gaia} DR3 parameter space, we identify 830 and 644 probable members of NGC~869 and NGC~884, respectively, with membership probabilities $P>0.75$. The resulting membership solution is stable for $m_{\rm clSize}=40$--100, indicating that the adopted samples are not sensitive to moderate changes in the clustering parameters. The radial density profiles further show that both clusters possess extended stellar structures, with NGC~869 being more spatially extended than NGC~884, with limiting radii of $80\pm1$ and $59\pm2$ arcmin, respectively.


The CCD $UBVRI$ photometry of the astrometrically selected members yields consistent astrophysical parameters for both clusters. We derive $E(B-V)=0.62\pm0.10$ and $0.60\pm0.08$ mag for NGC~869 and NGC~884, respectively, with evidence for differential reddening, particularly in NGC~869.
Adopting $[\mathrm{Fe/H}]=-0.30\pm0.10$ dex ($Z=0.008\pm0.002$), the CMD fits give a common age of $\sim12$ Myr and photometric distances of $\sim2.29$ kpc. The agreement between the CCD- and \textit{Gaia}-based fits, together with the sensitivity tests, indicates that these parameters are robust within the explored ranges.


The internal residual kinematics indicate that NGC~869 has a comparatively kinematically stable core, whereas NGC~884 shows a predominance of outward motions even in its central regions. The median projected residual velocity of NGC~869 increases rapidly beyond $\sim10$~pc, while that of NGC~884 is systematically higher and rises monotonically with radius. The outward velocities in NGC~884 nevertheless remain below $1~{\rm km\,s^{-1}}$ out to its outskirts, indicating a gradual rather than rapid expansion. This behaviour is consistent with the long-term dissolution inferred from the $N$-body simulations (see Figure~\ref{fig:dissolution}), in which NGC~884 loses its bound population more slowly than NGC~869. However, the dissolution timescales are model-dependent because the simulations assume initially virialised clusters. Given the observed expansion, this likely favours longer survival times, so the inferred timescales should be regarded as conservative upper estimates rather than precise predictions. 
 
We have conducted a comprehensive analysis of the dynamical state and long--term evolution of the Double Cluster by combining our best-fit astrophysical parameters, completeness--corrected mass estimates, and a suite of dynamical experiments. The results indicate that NGC 869 and NGC 884 do not form a gravitationally bound binary system. While the robust observational conclusion-supported by kinematics and binding analysis is that NGC~869 and NGC~884 do not form a gravitationally bound system at present, we also explored their forward evolution via dynamical models. Both the direct $N$-body simulations based on literature masses and the Monte Carlo two-particle experiments employing our corrected masses show a rapid increase in the cluster separation during the first 100 Myr, followed by a continued secular growth. 
In all cases, the separation increases beyond 1 kpc, indicating that the pair does not remain a long-lived gravitationally bound system.


The qualitative outcome is robust against uncertainties in the total masses. Although the completeness-corrected masses imply a significantly larger particle number for a full $N$-body simulation, the simplified two-particle approach provides a useful test of the sensitivity of the dynamical evolution to the adopted total masses. The observed, literature, and completeness-corrected mass estimates all produce the same qualitative outcome of continued separation, supporting the robustness of the conclusion.


The present-day dynamical state is consistent with the Double Cluster being a young but currently unbound pair. 
The binding analysis indicates that the present-day relative velocity is unlikely to be compatible with a gravitationally bound configuration.
NGC 869 and NGC 884 have been discussed as members of the young LISCA I cluster group \citep{dalessandro21}, providing a plausible common star-forming environment. Most such primordial groups tend to disaggregate~\citep{juan22}, providing a possible context for the present-day unbound state of the Double Cluster. However, these simulations constrain the evolution of the system from its present-day state and cannot determine whether the two clusters were gravitationally bound in the past.
	

In summary, current dynamical conditions indicate that the Double Cluster does not constitute a long‑lived gravitationally bound binary. The forward integrations show that the two clusters will continue to diverge under the influence of their relative motion and the Galactic potential.

\section*{Acknowledgements}

Part of the data are based upon observations carried out at the Observatorio Astronómico Nacional on the Sierra San Pedro Mártir (OAN-SPM), Baja California, México.  This paper has made use of results from the European Space Agency (ESA) space mission Gaia, the data from which were processed by the \textit{Gaia} Data Processing and Analysis Consortium (DPAC). Funding for the DPAC has been provided by national institutions, in particular the institutions participating in the \textit{Gaia} Multilateral Agreement. The \textit{Gaia} mission website is \url{http://www.cosmos.esa.int/gaia}. This paper has also made use of the WEBDA database, operated by the Department of Theoretical Physics and Astrophysics of the Masaryk University, Brno (the Vienna site is just a backup). This publication also uses the SIMBAD database and VizieR (\url{http://vizier.u-strasbg.fr/viz-bin/VizieR?-source=II/246}). This research was performed using services/resources provided by Grid UNAM, which is a collaborative effort driven by DGTIC and the research institutes of Astronomy, Nuclear Science and Atmosphere Science and Climate Change at UNAM in Mexico. We sincerely thank the Editor and the anonymous Reviewers for their careful and constructive comments, which helped us improve the clarity and quality of the manuscript.
Z.A. acknowledges financial support from the \.{I}lim Yayma Vakf{\i} (\.{I}YV).


\section*{Data Availability}

The photometric data of NGC~869 and NGC~884 can be requested from Raul Michel (rmm@astro.unam.mx).

\bibliographystyle{mnras}
\bibliography{references}



\label{lastpage}
\end{document}